\documentclass[fleqn,10pt]{wlscirep}
\usepackage[utf8]{inputenc}
\usepackage{graphicx}
\usepackage{tikz}
\usepackage[compat=1.0.0]{tikz-feynman}
\usepackage{amsmath,amsfonts,amssymb,verbatim,float}
\usepackage{ mathrsfs }
\usepackage{feynmf}
\usepackage{url}

\usepackage{caption}
\usepackage{subcaption}

\newcommand{\ER}{Erd\H{o}s-R\'enyi }

\title{Causal motifs and existence of endogenous cascades in directed networks with application to company defaults}

\author[1]{Irena Barja\v{s}i\'{c}}
\author[2]{Hrvoje \v{S}tefan\v{c}i\'{c}}
\author[3,4,+]{Vedrana Pribi\v{c}evi\'{c}}
\author[5,*,+]{Vinko Zlati\'{c}}
\affil[1]{Faculty of Science, University of Zagreb, 10000 Zagreb, Croatia}
\affil[2]{Catholic University of Croatia, Ilica 242,10000 Zagreb, Croatia}
\affil[3]{Zagreb School of Economics and Management, Zagreb, Croatia}

\affil[4]{Faculty of Economics, University of Ljubljana, Ljubljana, Slovenia}
\affil[5]{Division of Theoretical Physics, Rudjer Bo\v{s}kovi\'c Institute, Zagreb, Croatia}

\affil[*]{vinko.zlatic@irb.hr}

\affil[+]{these authors contributed equally to this work}

\begin{abstract}
Motivated by the detection of cascades of defaults in economy, we developed a detection framework for an endogenous spreading based on causal motifs we define in this paper. We assume that the change of state of a vertex can be triggered by an endogenous or an exogenous event, that the underlying network is directed and that times when vertices changed their states are available. In addition to the data of company defaults, we also simulate cascades driven by different stochastic processes on different synthetic networks. We show that some of the smallest motifs can robustly detect endogenous spreading events. Finally, we apply the method to the data of defaults of Croatian companies and observe the time window in which an endogenous cascade was likely happening. 
\end{abstract}
\begin{document}

\flushbottom
\maketitle
%
%
\thispagestyle{empty}

\section*{Introduction}

Dynamical processes in complex systems are most ubiquitously modeled using stochastic cellular automata and their continuous time analogue, interacting particle systems. These stochastic models are employed in descriptions of epidemics \cite{anderson1979population,anderson1992infectious,kiss2017mathematics}, systemic risk in financial industry \cite{battiston2012debtrank,gai2010contagion}, propagation of information in society \cite{weng2012competition}, queuing \cite{kendall1953stochastic} etc. In some of these systems, dynamics can lead to the formation of cascades like, for example, a fast evolving contagion or a series of company defaults. Detecting such events can be very important as is the case in new previously unknown pathogen or for understanding what drives some default avalanche in economy.

However, merely the possibility of a type of a contact process in the class of complex systems of interest does not imply that temporal data collected from some such system should be modeled using the aforementioned stochastic processes.
The lack of contact data, necessary to reconstruct the network to model the spreading processes (pathways a new pathogen takes \cite{volz2007susceptible,blower2011importance} or the reasons for companies default \cite{vodenska2019interdependence}); the lack of some layers in a multiplex network that would most precisely represent the real system \cite{mucha2010community,boccaletti2014structure,bookMoreno2018}, or simply no network contribution in the spreading - are some of the instances in which modeling of the process is best done as a non-network field effect.

Therefore, the question we are addressing in this work is, whether the observed dynamics in the data can in a justified manner be explained through stochastic cellular automata on the available network - therefore being  endogenous with respect to the data, or it would be more appropriate to model it as an purely exogenous (field effect) influence.

 The motivation for this research is directly related to economics, more concretely - a problem of default cascades \cite{battiston2012default, brummitt2015cascades}.
Knowing what causes companies to fail to repay its debt, i.e. go into default, is crucial for every economy. Sometimes it can be related mainly to exogenous events like for instance loss of market access, loss of access to finance, monetary shocks etc. In other cases the defaults in the system (network) are mostly endogenous and are related to inability of different actors (vertices) to fulfill their obligation to their lenders. This type of events is common in financial systems that are not well regulated and it is believed to be a significant contributor to largest systemic events \cite{battiston2012default, brummitt2015cascades,zlatic2015reduction}  even in the well regulated systems.

Financial contagion on networks is usually modeled using balance sheets of nodes \cite{gai2010contagion}, where different categories in the balance sheet insulate against financial distress. Kobayashi \emph{et al.} in \cite{kobayashi2014efficient} illustrates the equivalence of the balance sheet approach with the threshold model of cascades presented in \cite{watts2002simple}. In \cite{acemoglu2015systemic} Acemoglu et al. explore network architecture contributing to or insulating against financial contagion between banks. 
On the other hand, very few papers exist that study transmission of financial shock between firms that utilize entire networks. Contagion is mostly studied through the lens of credit chains firstly outlined in \cite{kiyotaki1998credit} as important channels for propagation of financial distress. 
Defaults on trade credit have been identified as a major cause of firms distress, as explicitly modelled in both \cite{boissay2006credit} on US data and \cite{jacobson2015trade} on Swedish data. 
Acemoglu et.al. in \cite{acemoglu2012network}, as well as \cite{hertzel2008inter} and \cite{oberfield2018theory}  aim to develop a network theory of production which would meaningfully model links between firms. 
Others, such as \cite{gao2015effects} construct a business network from supply chain data. 
Contagion within production network is explored in \cite{battiston2007credit} where two types of bankruptcies exist: either by a random shock to revenues or costs or when a creditor is not paid by debtor. 
However, a way to discern between these two types of defaults has not yet been developed. 
Economic motivation for this paper is to develop a basis for an empirical strategy which would allow us to distinguish from the data the existence of shocks transmitted from other nodes as opposed to field effects that affect node failure.

Here we investigate the methodology for detection of endogenous propagation in financial networks, which was previously used in the investigation of Croatian company defaults without theoretical justification \cite{Zlatic}. There, the problem was treated using randomized reference models (RRM) \cite{gauvin2018randomized} of a directed temporal network with the companies and their times of default on vertices, and their mutual debts on edges, pointed from borrower to lender. In addition to single causal edges used there, we introduce \emph{causal motifs} with two and three edges and the largest component of causal edges as variables we use for the test statistics, but we omit any other data like debt (edge weights) in order to make the method usable in as many situations as possible. This methodology should in principle be applicable to any spreading phenomena (with two states) on any directed network. 

RRM were also used in temporal embedding \cite{torricelli2020weg2vec}, inference of structures in communication networks \cite{lehmann2019fundamental} and analysing collective behaviour in social networks \cite{mellor2018analysing}.

Network motifs have a long tradition of being used as a tool for inference in complex networks \cite{milo2002network}. They are defined as small subgraphs that can be observed with different frequencies in the data. They were previously used to understand metabolic and other biological networks \cite{milo2002network,shen2002network,alon2007network}, the properties of ecological system through food webs \cite{ristl2014complex}, in economical setting to understand corporate governance \cite{ohnishi2010network,takes2018multiplex}, and organization of knowledge in Wikipedia \cite{zlatic2006wikipedias}. Motif analysis was also efficiently extend to temporal networks \cite{paranjape2017motifs}. 

Another line of research relevant for this work is the inference of external and internal influence on the propagation of information. Previous works were mainly embedded in social networks and questioned if the influence of peers in a network is dominant to outside influences or not
\cite{anagnostopoulos2008influence,pivskorec2017modeling,myers2012information}. The direction of this line of research is mostly related to the estimation of parameters to proposed models of information spreading.

In order to validate our methodology, besides using the original data of company defaults, we create artificial data with controlled parameters. Using Kolmogorov-Smirnov test, z-score and Mahalanobis distance, we determine which type of motifs is the most successful test statistic. For convenience, all the scripts used in this paper and the company default data are available in our Git-Hub repository of in the Supplementary information \cite{GIT-HUB}. Since the parameter space needed for the evaluation of all the possible processes and networks is enormous we focus on 2 types of processes on relatively small networks for which we expected the results to be the least conclusive, therefore providing a lower bound for the significance of the results. 

\subsection*{Data on defaults}
The data is collected from the Croatian Financial Agency website which publicly discloses all documents related to the Chapter 11 type bankruptcies which involved debt renegotiation and restructuring. A new law was passed at the end of 2012 which defined the criteria companies had to meet in order to file for this type of bankruptcy: failing to attain liquidity over the course of 60 days and at most 21 days passing since their insolvency onset. The companies are obliged by law to list an extensive list of its creditors on the day they file for the procedure. However, the format of public data required extensive data mining and a big effort in cleaning these data to produce inputs which could be used in network formation.

The orderly data consists of a list of debtors, an exhaustive list of their creditors, the starting time and the length of the pre-bankruptcy settlement and the amount of debt per creditor. The data spans from December 19th 2012 to February 26th 2014, and contains 25469 vertices and 52507 edges, where debtors are exclusively firms and creditors range from banks, private and public firms, government and individuals. Total amount of reported debt was 5.97 billion euros, which corresponded to 13,6 percent of Croatia's GDP in 2014. The choice of specific time frame for dataset was intentional; in order to observe cascades, data collection began when the largest and most interconnected debtors filed for the procedure soon after the law was introduced at the end of 2012, following the initial illiquidity build-up, while number and size of firms significantly decreased in the next two years. Additionally, we only include recession years to avoid business cycle effects of economic recovery which may exogenously affect probabilities of defaults. i.e. may exert a different field effect on our network.

\begin{figure*}[ht]
\centering
\includegraphics[width=1\linewidth]{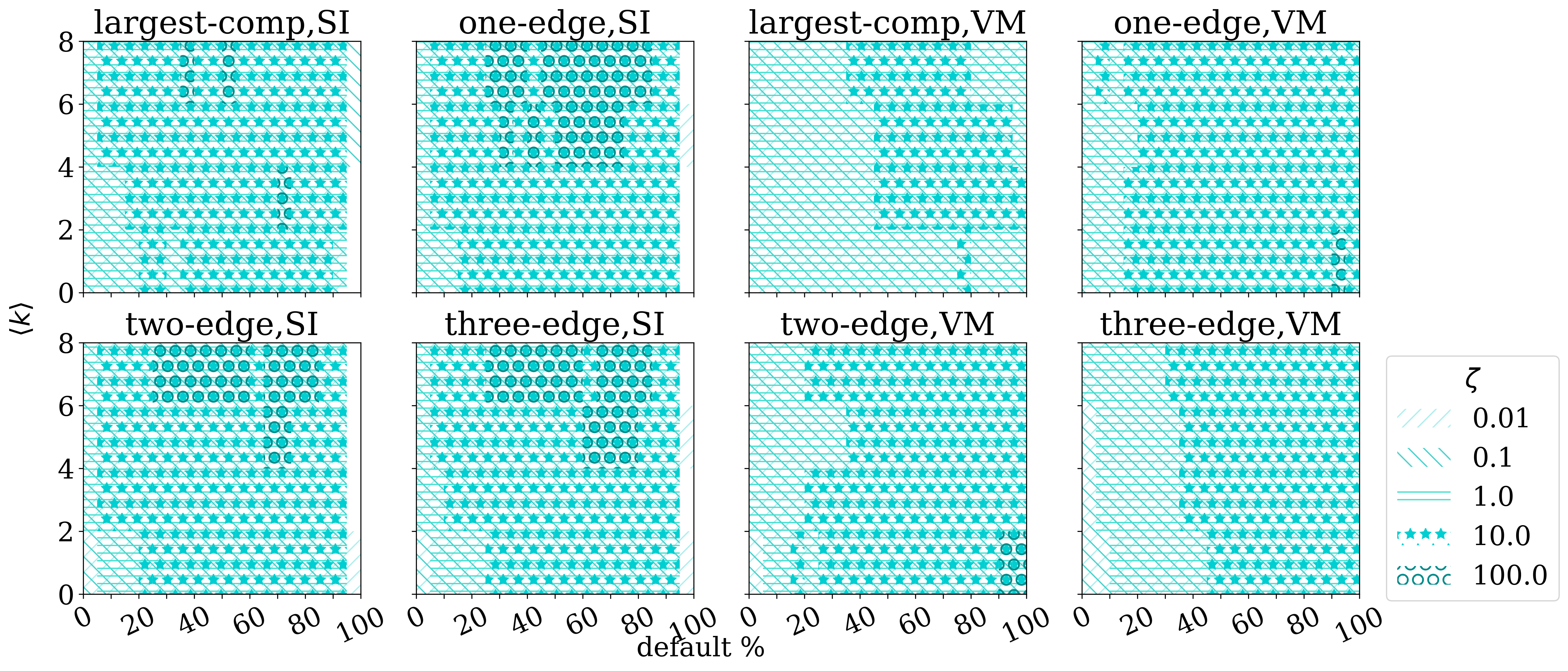}  
\caption{KS-test results for SI and VM processes on a network of $N=1000$ vertices. 
The significant statistic $(p < 0.01)$ for a given percentage of default, $\zeta$ is marked with the corresponding pattern. On x-axis we show the percentage of defaults in the network and on y-axis the average degree of the network.}
\label{fig:KS_SI}
\end{figure*}

\section*{Results}
The final form of the results, after generating $N_{graph}=100$ different graphs, simulating $N_{process}=10$ processes on every graph, and creating $N_{shuffle}=100$ randomized reference models for each process, are two distributions of the statistic counts. One of them is the distribution of a statistic recorded on the original process, and the other the distribution of the statistic on the networks that result from time-shuffling. 

To quantify the difference between the distributions, we use the two-sample Kolmogorov-Smirnov test. For a more precise comparison between one instance of a process and its null distribution, we employ the z-score. Finally, to distinguish between the contributions of individual motifs within a class of motifs, we use the Mahalanobis distance, which is a generalization of the z-score.

\begin{figure*}[ht]
\begin{subfigure}{.33\textwidth}
  \centering
  \includegraphics[width=1\linewidth]{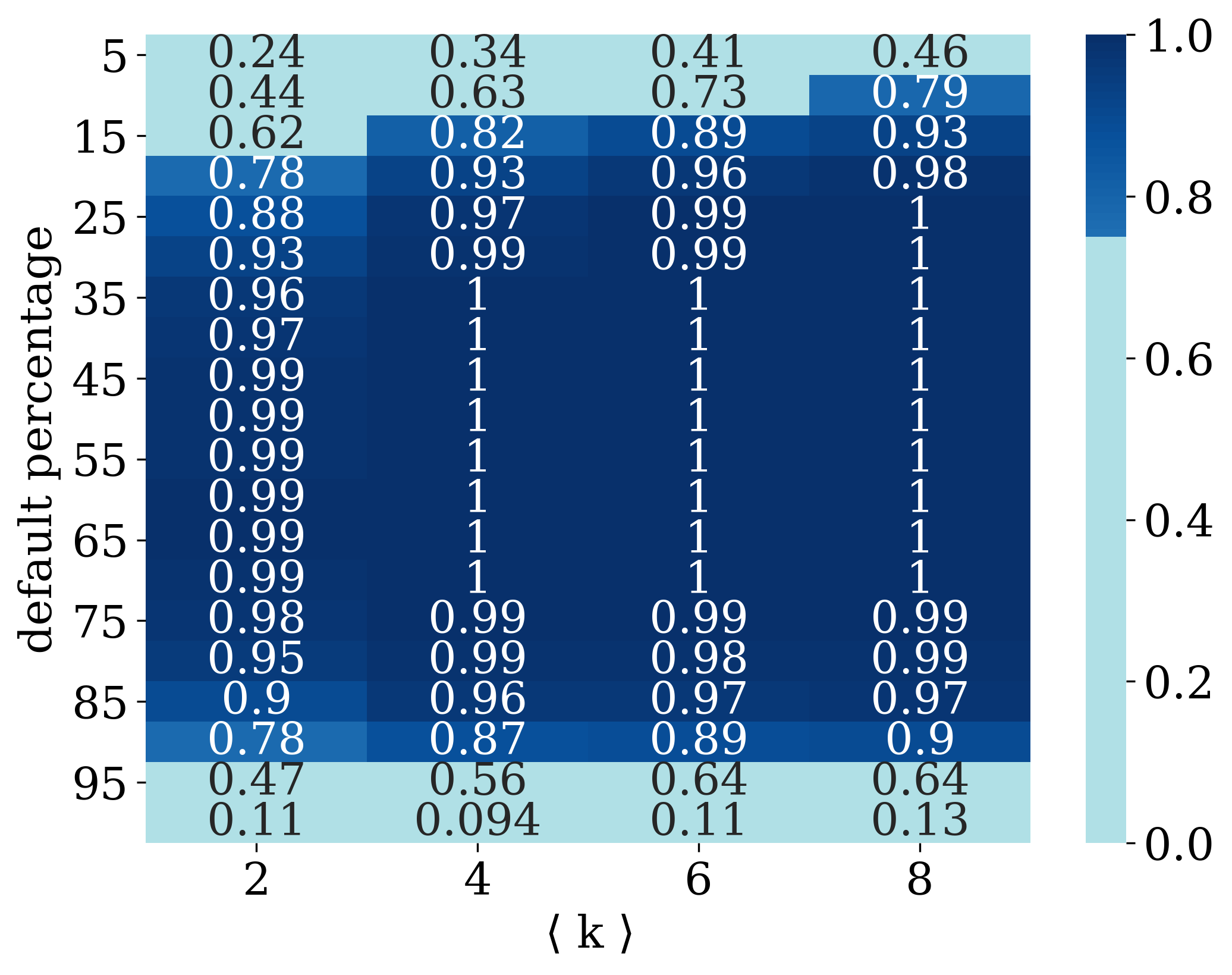}  
  
  \caption{}
  \label{fig:SI-results-a}
\end{subfigure}%
\begin{subfigure}{.33\textwidth}
  \centering
  \includegraphics[width=1\linewidth]{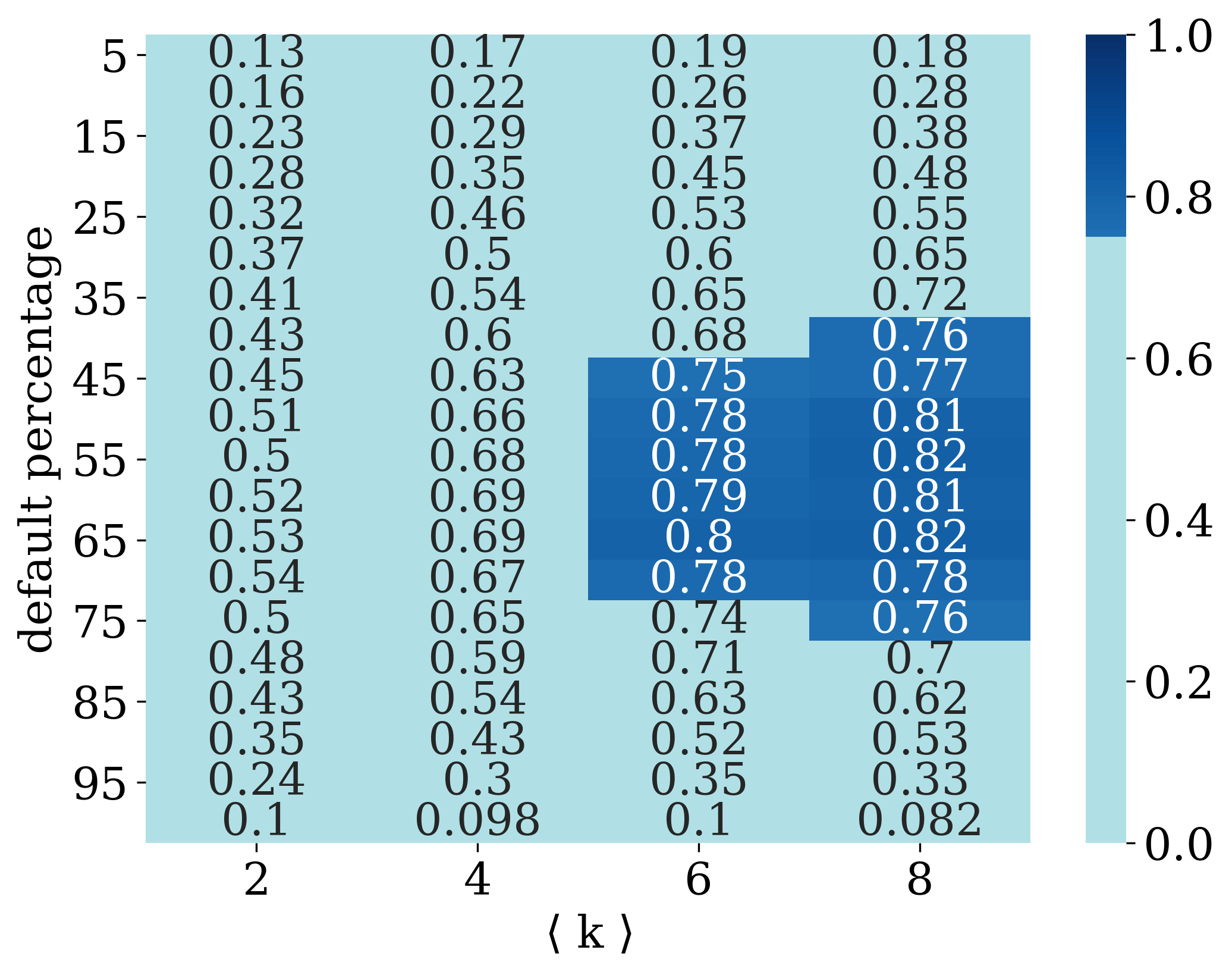}  
  
  \caption{}
  \label{fig:SI-results-b}
\end{subfigure}
\begin{subfigure}{.33\textwidth}
  \centering
  \includegraphics[width=1\linewidth]{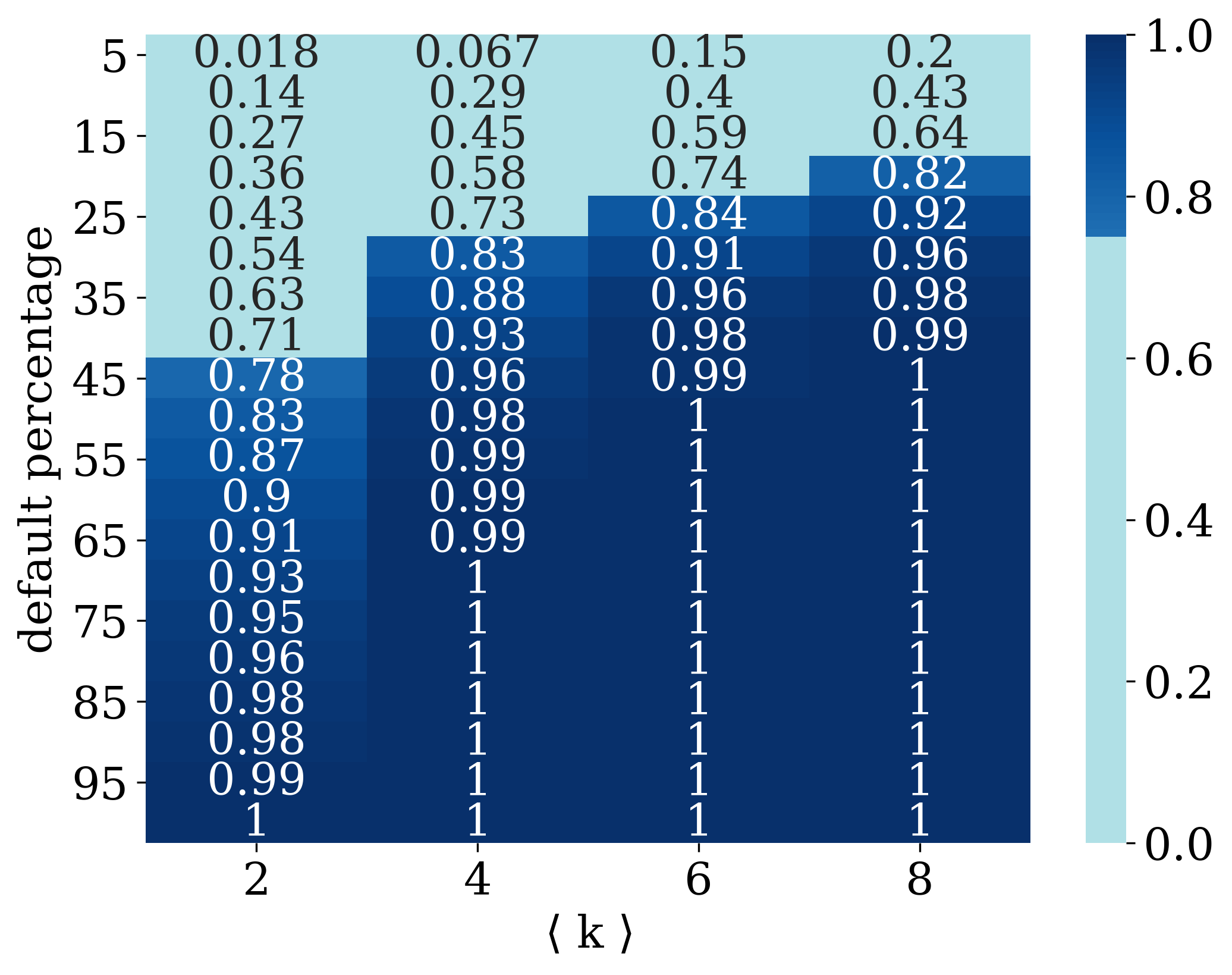}  
  
  \caption{}
  \label{fig:SI-results-c}
\end{subfigure}

\begin{subfigure}{.33\textwidth}
  \centering
  \includegraphics[width=1\linewidth]{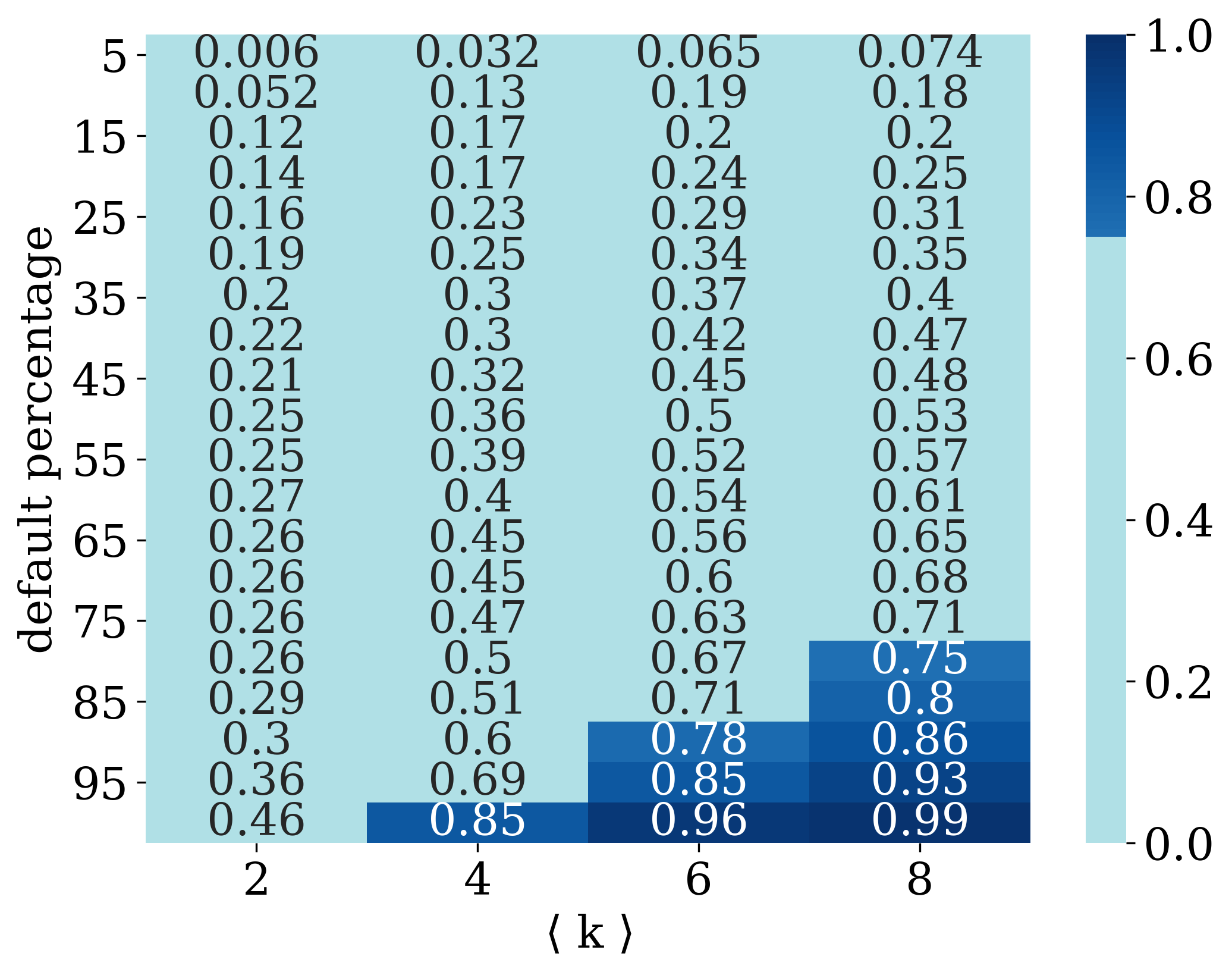}  
  
  \caption{}
  \label{fig:SI-results-d}
\end{subfigure}
\begin{subfigure}{0.33\textwidth}
  \centering
  \includegraphics[width=1\linewidth]{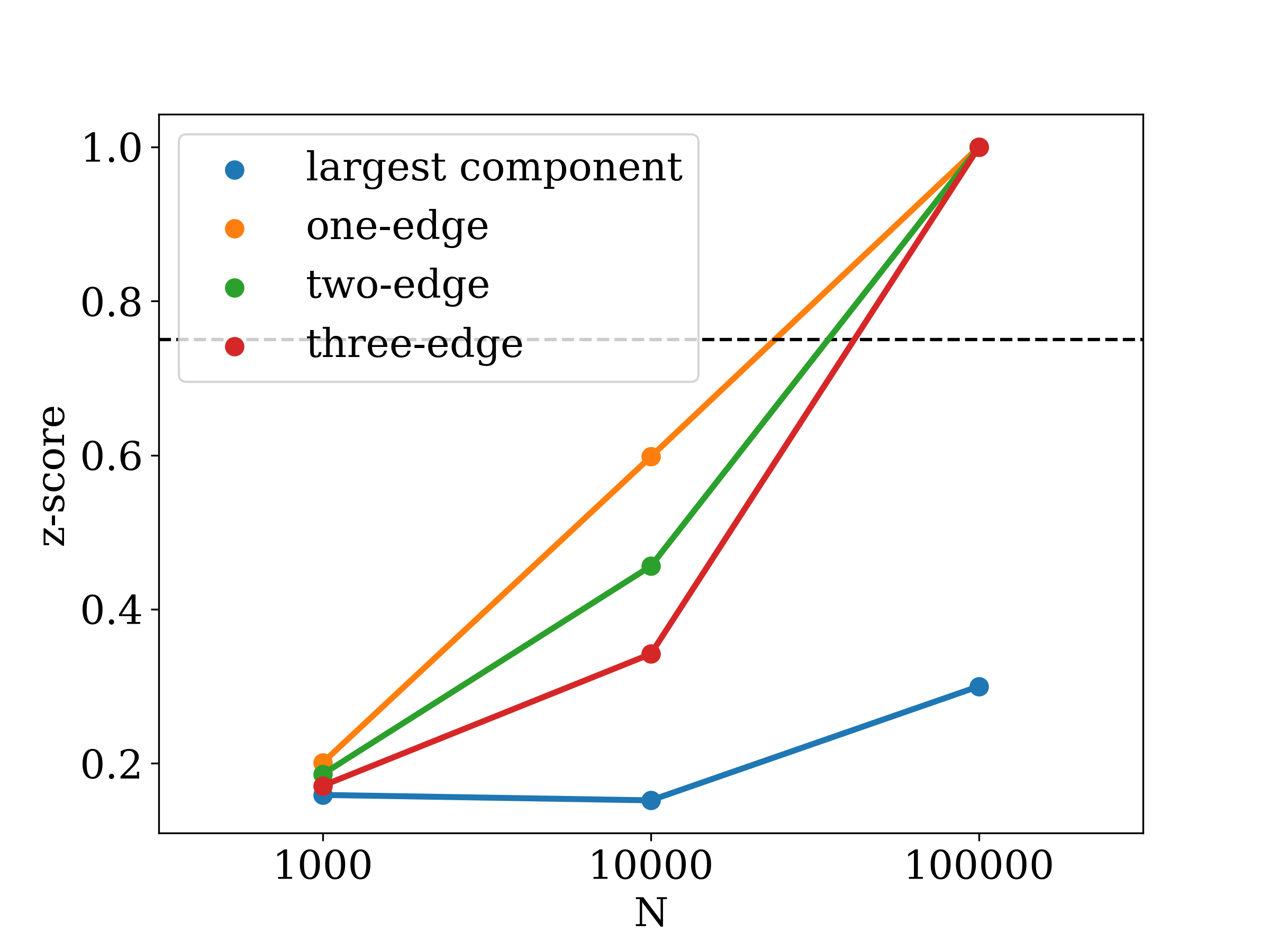}
  
  \caption{}
  \label{fig:SI-results-e}
\end{subfigure}
\begin{subfigure}{0.33\textwidth}
  \centering
  \includegraphics[width=1\linewidth]{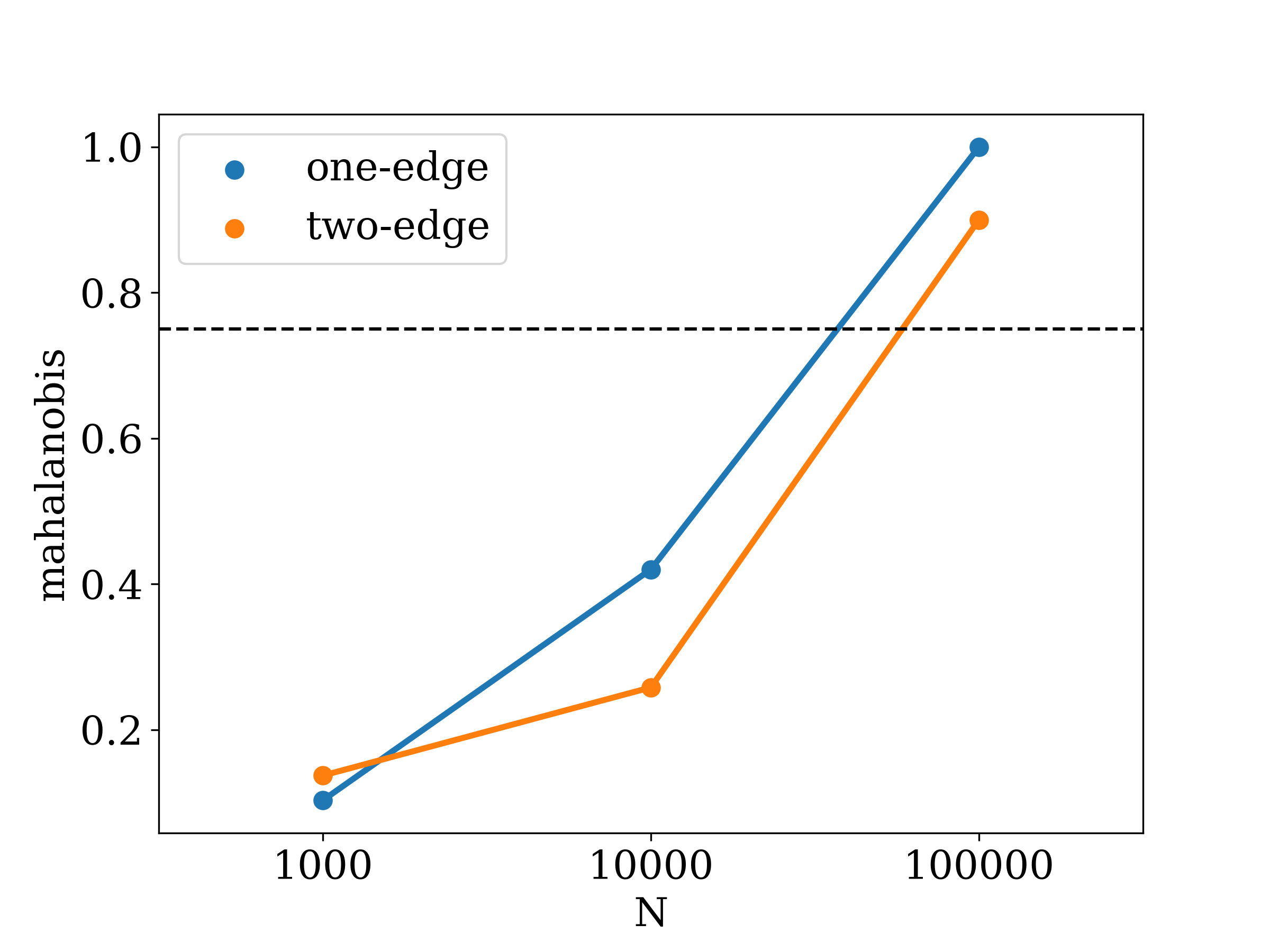}
  
  \caption{}
  \label{fig:SI-results-f}
\end{subfigure}
\caption{SI process results.
In panels a) and b) we present the fraction of one sided z-score results significant with $p< 0.1$ (dark blue), depending on the percentage of defaults in the network (y-axis) and average degree of the network (x-axis), for an SI process with $\zeta=1$ and $\zeta=4$, respectively. In panels c) and d) we present in the same form the results of the Mahalanobis distance of two-edge motifs. For these figures, the number of vertices equals $N=1000$, on each of 100 realized networks, 10 processes are simulated and 100 random shuffles are created. In panels e) and f) we fix the network default at $25 \%$ and present the percentage of significant tests depending on the size of the network for e)- z-score and f)-Mahalanobis distance with $\langle k\rangle=4$ and $\zeta=10$. Horizontal lines represent the case in which $75\%$ of simulations exhibit significant difference from the RRF-model.
}
\label{fig:SI-results}
\end{figure*}

\begin{figure*}[ht]

\begin{subfigure}{.32\textwidth}
  \centering
  \includegraphics[width=1\linewidth]{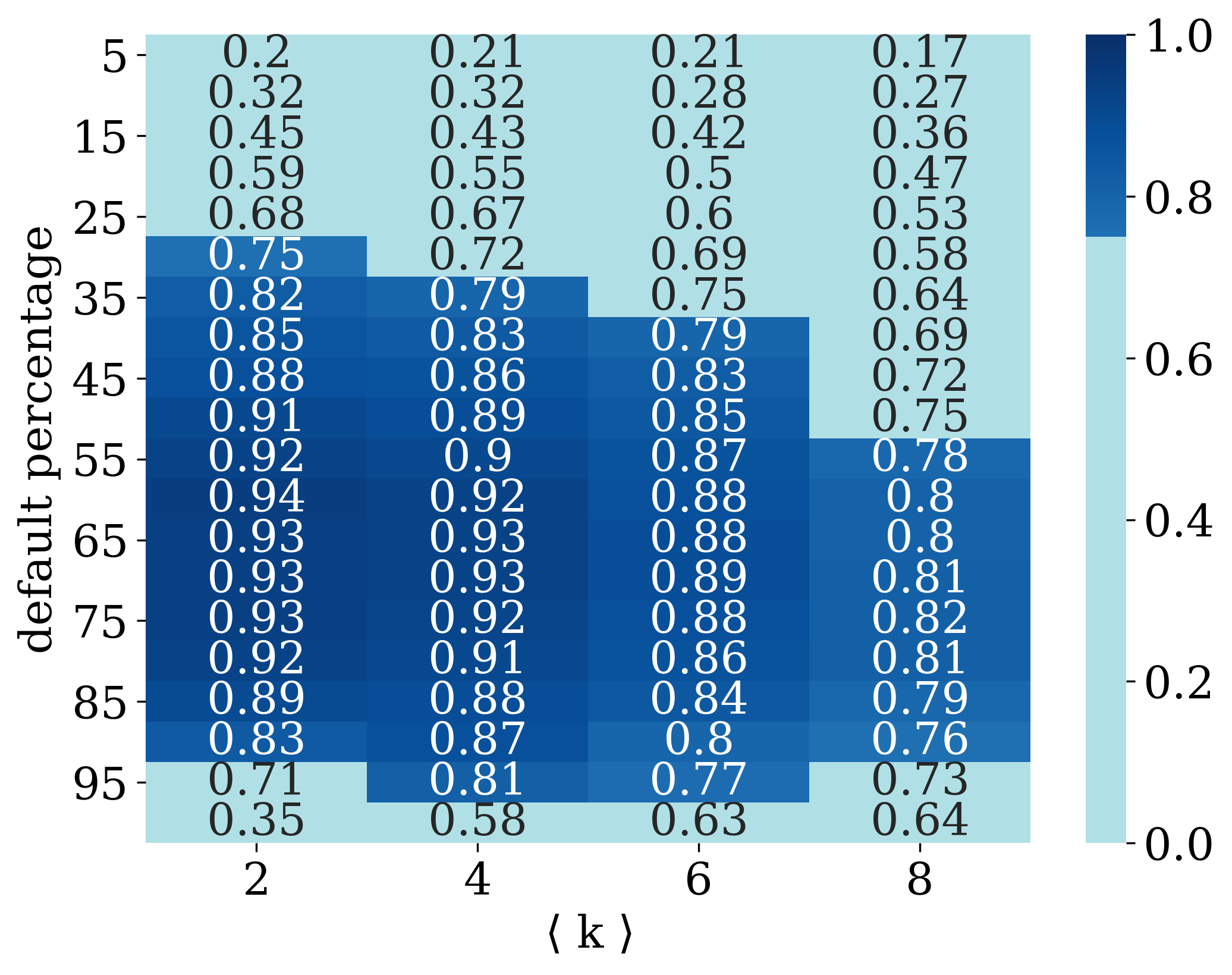}  

  \caption{}
    \label{fig:VMa}
  \end{subfigure}
\begin{subfigure}{.32\textwidth}
  \centering
  \includegraphics[width=1\linewidth]{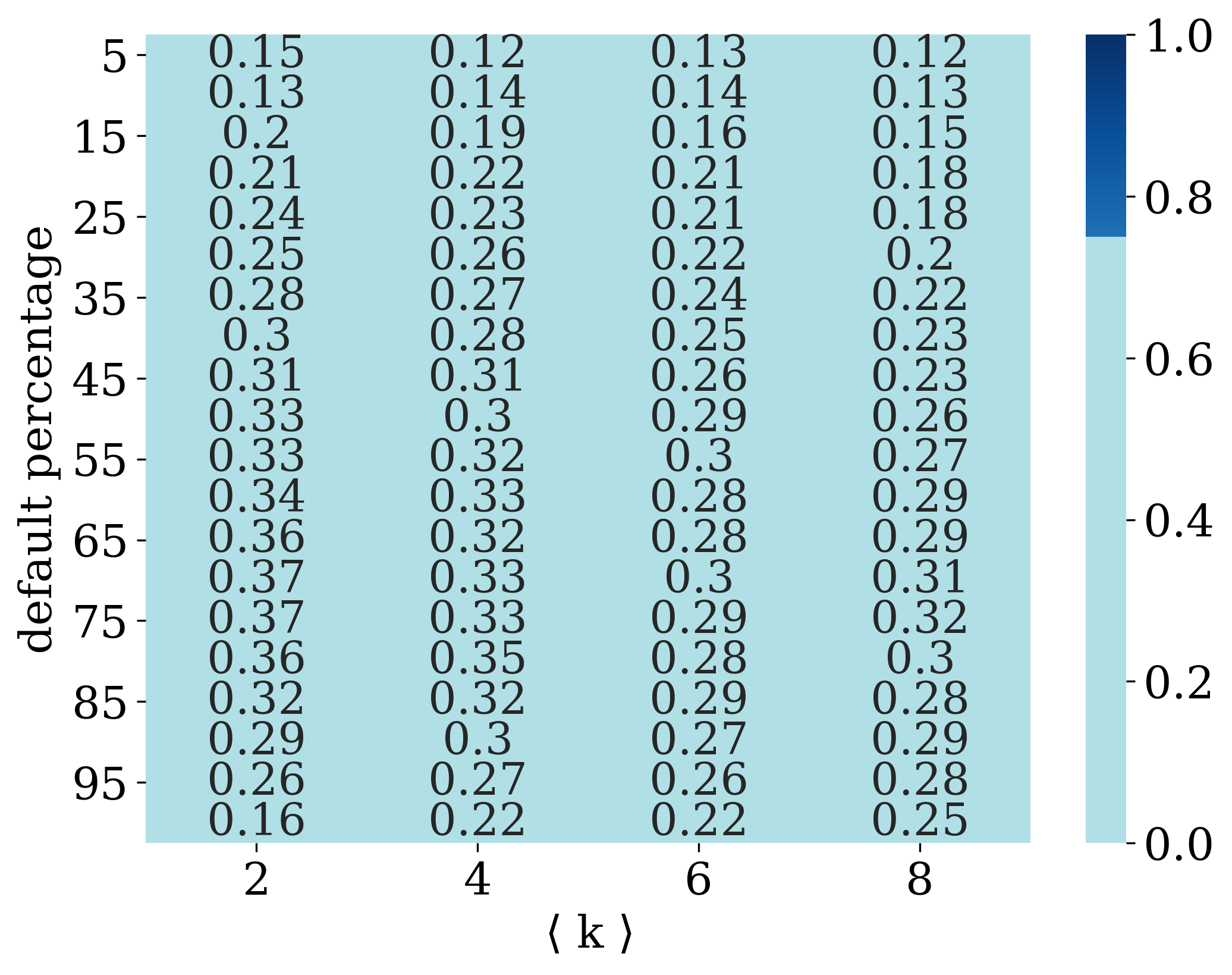}  
 
  \caption{}
  \label{fig:VMb}
\end{subfigure}
\begin{subfigure}{.32\textwidth}
    \centering
  \includegraphics[width=1\linewidth]{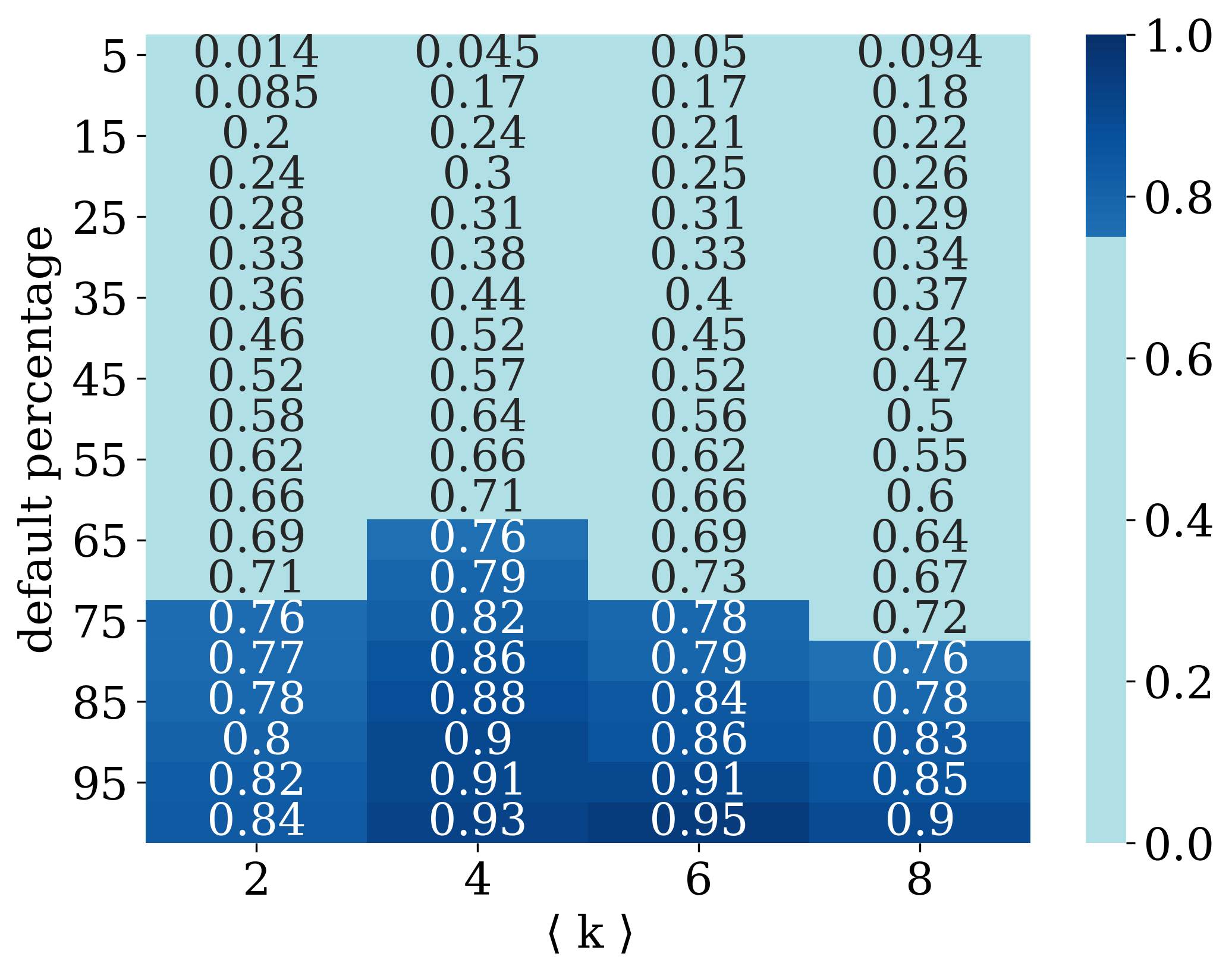}  
 
  \caption{}
  \label{fig:VMc}
\end{subfigure}

\begin{subfigure}{.32\textwidth}
  \centering
  \includegraphics[width=1\linewidth]{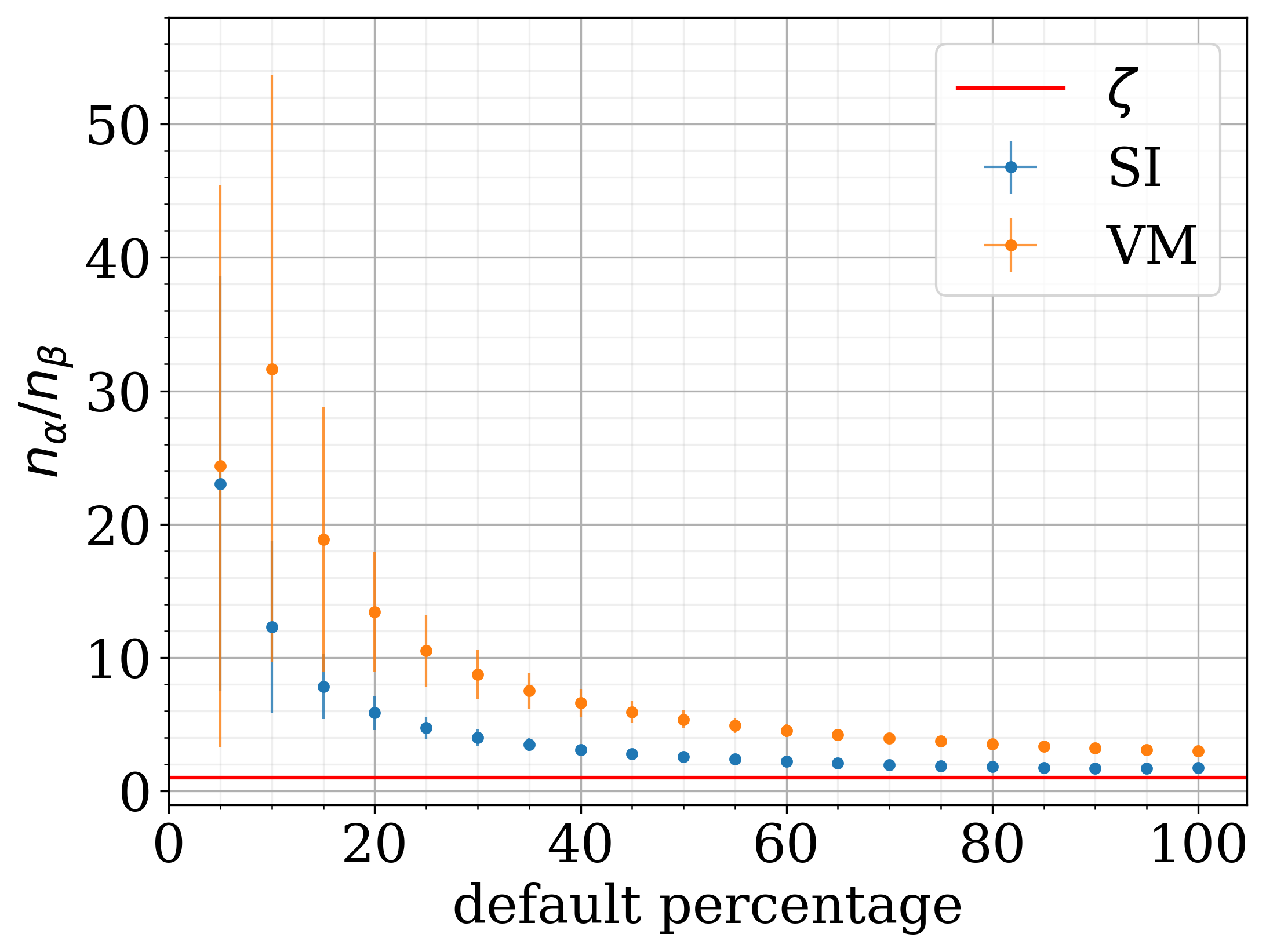}  
 
  \caption{}
  \label{fig:VMd}
\end{subfigure}
\begin{subfigure}{.32\textwidth}
  \centering
  \includegraphics[width=1\linewidth]{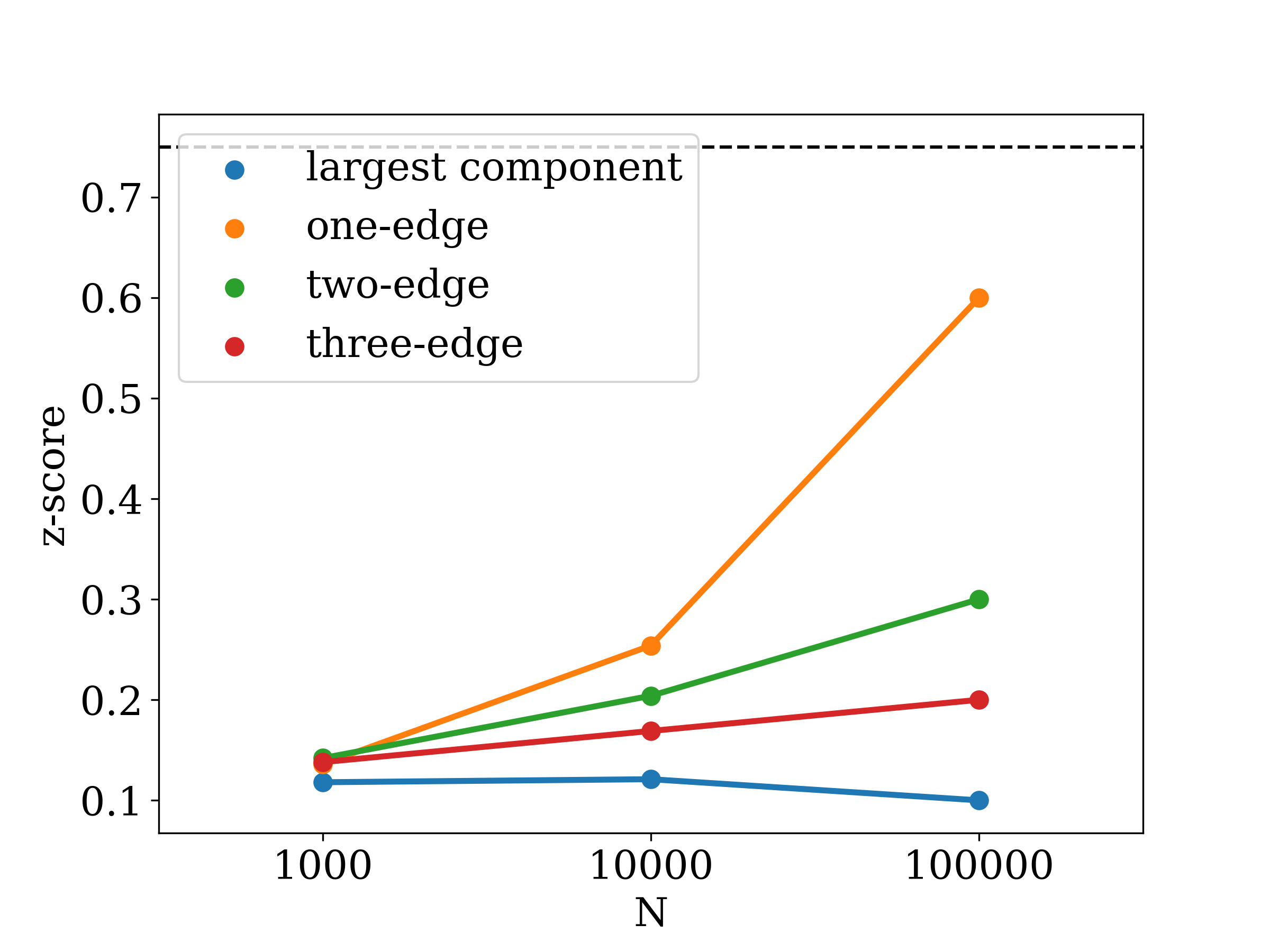}  
 
  \caption{}
  \label{fig:VMe}
\end{subfigure}%
\begin{subfigure}{.32\textwidth}
  \centering
  \includegraphics[width=1\linewidth]{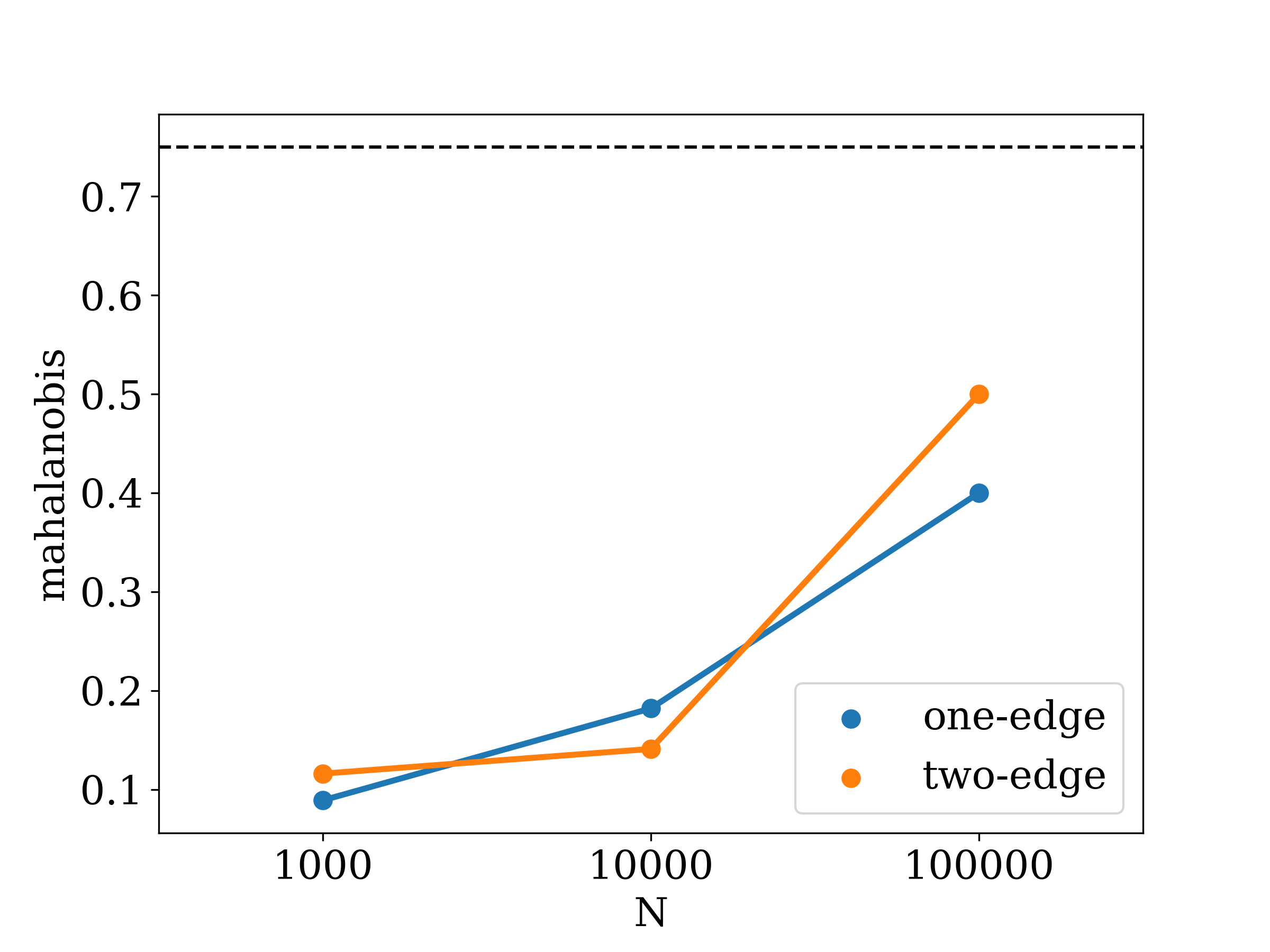}  

  \caption{}
  \label{fig:VMf}
\end{subfigure}
\caption{VM process results.
In panels a) and b) we present the fraction of one sided z-score results significant with $p< 0.1$ (dark blue), depending on the percentage of defaults in the network (y-axis) and average degree of the network (x-axis), for a VM process with $\zeta=1$ and $\zeta=4$, respectively. In panel c) we present the results in the same form the results of the Mahalanobis distance of two-edge motifs with $\zeta=1$. For these figures, the number of vertices equals $N=1000$, on each of 100 realized networks, 10 processes are simulated and 100 random shuffles are created. In panel d) we present the ratio of the number of nodes that defaulted through the exogenous process $n_{\alpha}$ and the endogenous process $n_{\beta}$, for different default percentages ($\langle k\rangle=4$ and , $\zeta=1$).  In panels e) and f) we fix the network default at $25 \%$ and present the percentage of significant tests depending on the size of the network for e)- z-score and f)-Mahalanobis distance with $\langle k\rangle=4$ and $\zeta=10$. Horizontal lines represent the case in which $75\%$ of simulations exhibit significant difference from the RRF-model. \cite{GIT-HUB}}
\label{fig:VM-results}
\end{figure*}  
In the simulated processes, we quantify the endogenous component of the process with the parameter $\alpha$ - which is equal to the probability per unit time that the vertex changes a state without any contributing effect from the network, and the exogenous component with the parameter $\beta$ - which is related to probability that the change of state came from an immediate neighbour. Parameter $\beta$ is process dependent and details are given in the section "Methods". The ratio $\zeta=\frac{\alpha}{\beta}$  between the rates of the two components is a control parameter we use to present results. For a single exogenous and a single endgenous process, it would represent the ratio of the changes of exogenously and endogenously defaulted vertices. However, as the exogenous default of every vertex is modeled with its own Poisson process with rate $\alpha$, and the endogenous default of vertices with defaulted neighbours can be caused by transmissions through those edges, each with rate $\beta$, the actual ratio ($\Delta n_{\alpha}/\Delta n_{\beta}$) will in general be different from $\zeta$. As can be seen in Supplement, it is related to the parameter $\zeta$, but also to the total number of vertices, the mean degree of the network and the percentage of the network default.

We focus on small networks consisting of $N=1000$ nodes (typical size of financial networks), they are small enough to allow for a comprehensive research and we later show that increasing the network size raises the significance of detection substantially. 

In the last section we show the results of applying the methodology to the real data of Croatian companies' defaults. 

\begin{figure*}[ht]
\centering
\begin{subfigure}{0.45\textwidth}
  \centering
  \includegraphics[width=.8\linewidth]{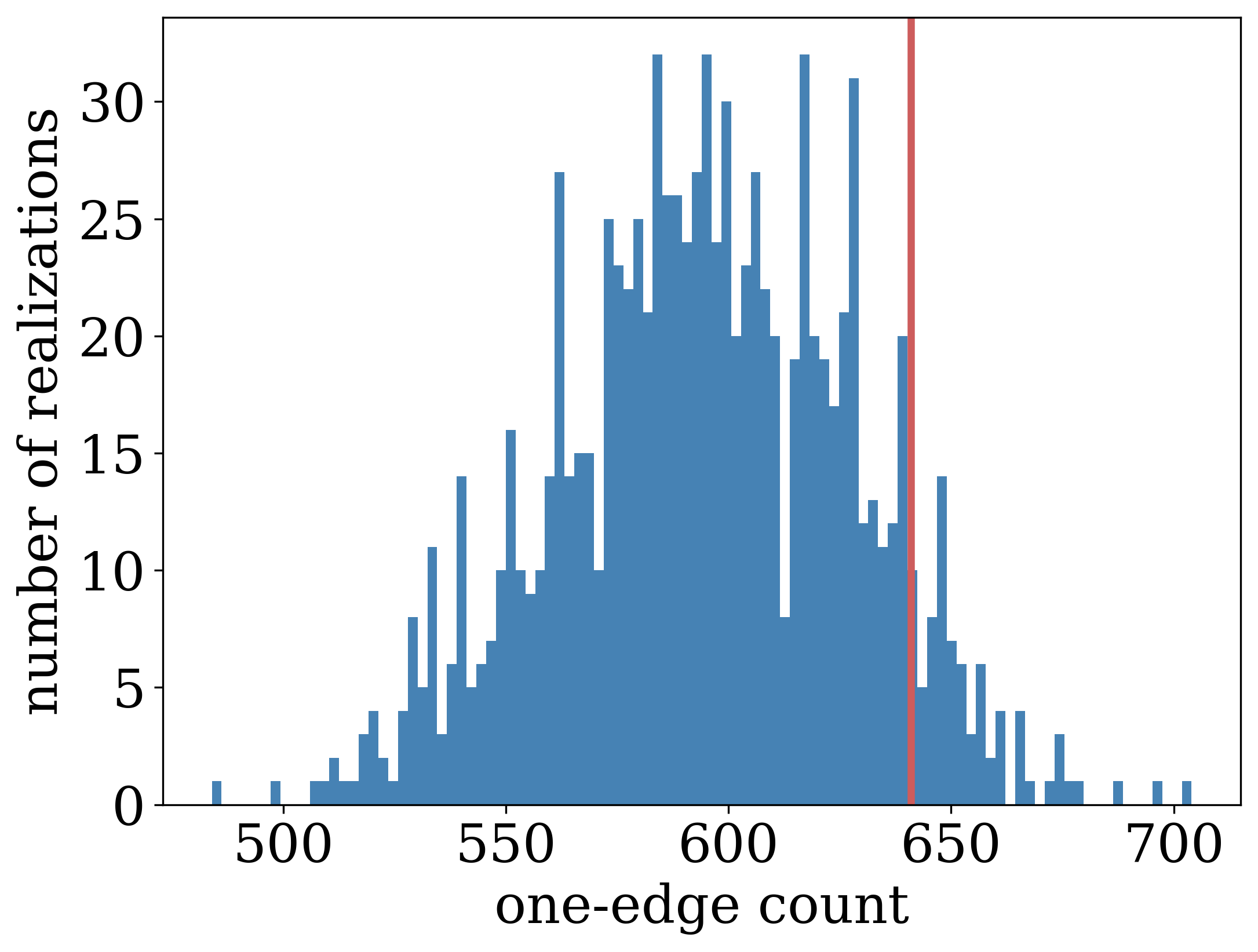}  
  \caption{}
  \label{fig:CroDefa}
\end{subfigure}
\begin{subfigure}{.45\textwidth}
  \centering
  \includegraphics[width=.8\linewidth]{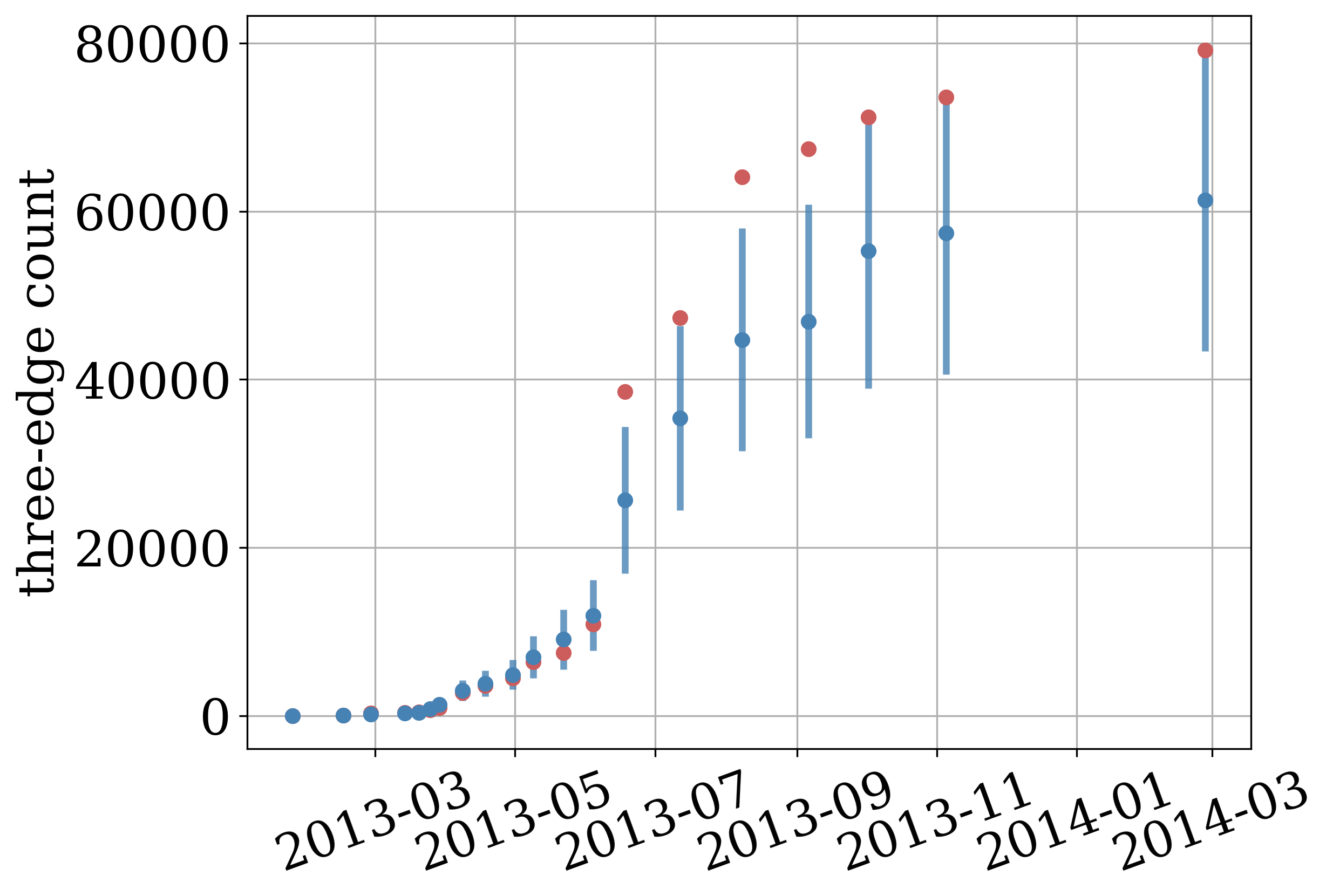}  
  \caption{}
  \label{fig:CroDefb}
\end{subfigure}
\\[1ex]
\begin{subfigure}{.24\textwidth}
  \centering
  \includegraphics[width=0.8\linewidth]{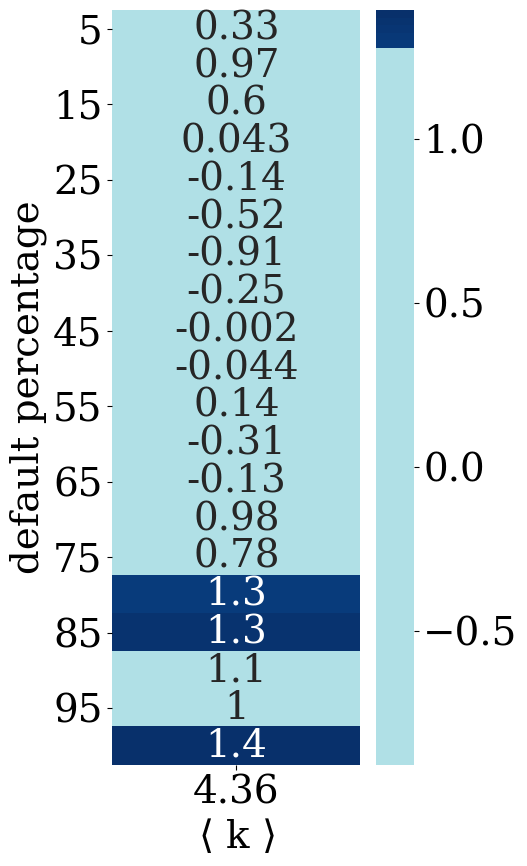} 
  \caption{}
  \label{fig:CroDefc}
\end{subfigure}
\begin{subfigure}{.24\textwidth}
  \centering
  \includegraphics[width=0.8\linewidth]{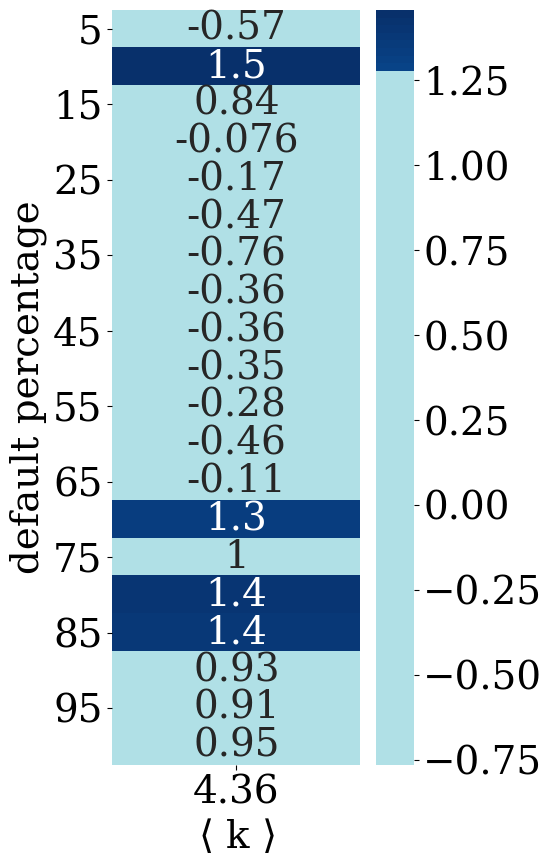}  
  \caption{}
  \label{fig:CroDefd}
\end{subfigure}%
\begin{subfigure}{.24\textwidth}
  \centering
  \includegraphics[width=0.8\linewidth]{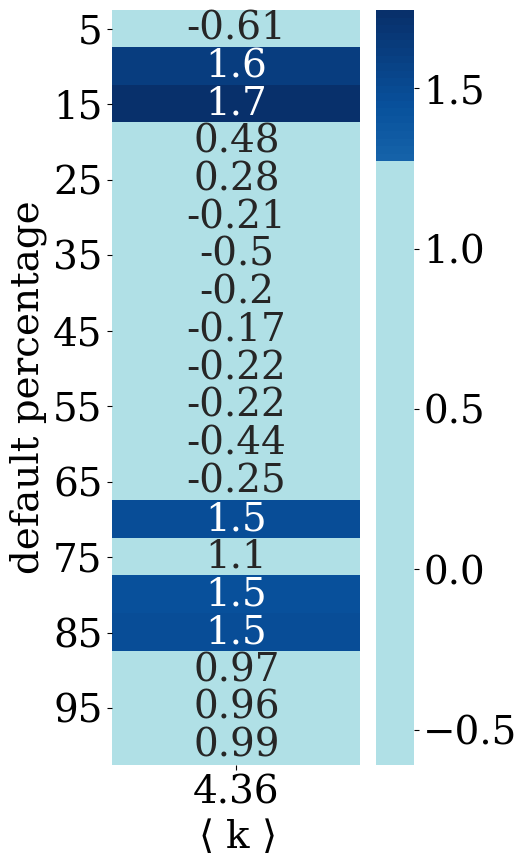}  
  \caption{}
  \label{fig:CroDefe}
\end{subfigure}
\begin{subfigure}{.24\textwidth}
  \centering
  \includegraphics[width=0.8\linewidth]{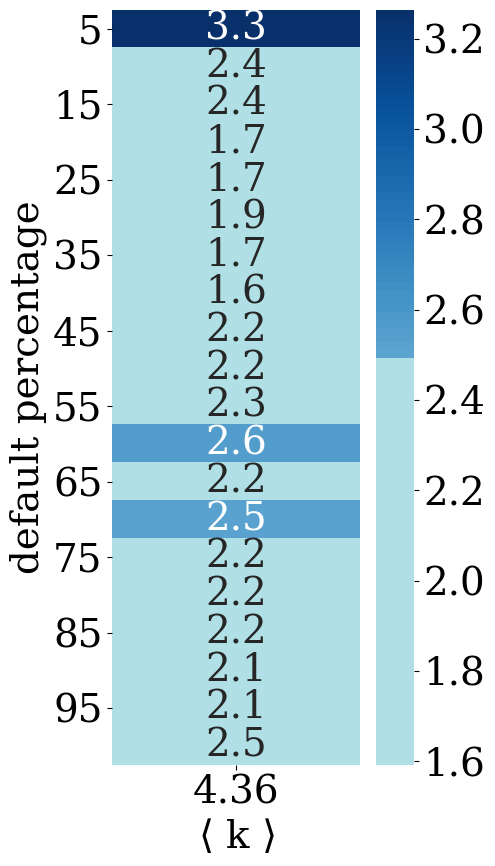}  
  \caption{}
  \label{fig:CroDeff}
\end{subfigure}
\caption{Defaults of Croatian companies results.
In figure a) we present the frequency of one-edge motifs in all the data, compared to a RRM (1000 shuffled instances) histogram of the same statistic. In panel b) we present the count of three-edge motifs (red points) compared to RRM realizations represented with blue points (mean) and a blue line (standard deviation) for different percentages of network default. In panels c) (one-edge) d) (two edge) and f) (three edge) are the z-scores of the test statistics on data presented in previous figure. In panel f) we show the two-edge Mahalanobis distance. Significant cases ($p < 0.1$) in the last four figures are colored dark blue.}
\label{fig:CroDef}
\end{figure*}

\subsection*{SI model}

 The first model of a spreading process we investigate, is a variant of an SI model. As described in Methods results are interpreted using the two sample Kolmogorov-Smirnov test (Fig. \ref{fig:KS_SI}), the z-score (Fig. \ref{fig:SI-results}
) and Mahalanobis distance (Fig. \ref{fig:VM-results}). %

The distribution of simulated processes is compared to the respective randomized distributions, using KS-test, to infer if the distributions of realizations are different. If they are not, the individual realization can not be expected to give meaningful statistically significant signal. From Fig. \ref{fig:KS_SI}, four left panels, we can see that the distributions of all investigated motifs are significantly different from our null model except for the cases when almost the entire network is defaulted and the cases when the number of defaults is low. The samples with the largest values of $\zeta$ ($\zeta = 100$) are significantly different for larger mean degrees and an intermediate range of the network default. For the largest values of $\zeta$, one, two and three edge motifs provide significant difference in a bit larger part of the parameter space than the largest component.

Addressing the one-sided z-score results \ref{fig:SI-results}, one-edge motif statistics prevails for all simulated $\zeta$ parameters if the network degree $\langle k \rangle$ is 4 or less. Causal motif statistics always outperforms largest component statistics, by a large margin. Surprisingly, causal motifs of higher order with two and three edges show less significant results than one edge statistics, because even though they separate first moments of the two distributions better, they also introduce a larger variance. In Figs.\ref{fig:SI-results-a} and \ref{fig:SI-results-b} we present a fraction of simulated processes that were significantly different ($p<0.1$) from the null-model, and one can see how the increase of $\zeta$ shrinks the area of satisfactory results. Unsurprisingly, with the increase of the system size we are able to find that our method can distinguish the existence of an endogenous process with statistical significance even in these extreme cases  Fig.\ref{fig:SI-results-e}.

The causal motif test statistics on both the original and randomized networks converge on average in time to a constant fraction of the total number of possible motifs. For one-edge motifs that fraction is $1/2$ and for two-edge motifs $1/4$ of their total number. On randomized networks, where time ordering of the process is completely destroyed, that kind of convergence is expected. Nevertheless, even with a strong endogenous driving, average numbers of causal motifs formed by an SI process converge in time to the same limits. If we observe a completely endogenous process of defaults until all vertices on the \ER network default, we can calculate the final number of causal one-edge motifs by summing the expected number of causal one-edge motifs in each step. The probability that one of the $\langle k_{out} \rangle$ outgoing edges reaches a susceptible vertex equals to the fraction of susceptible vertices $1-t/N$, since the \ER network is homogeneous and the SI process treats all out-neighbours as equal. We see then that the number of causal one edges is:
\begin{align}
    C^{(1)} = \langle k_{out} \rangle \sum_{t = 1}^N \frac{N-t}{N} &= \frac{\langle k_{out} \rangle}{N} \frac{N^2-N}{2} \approx \frac{N \langle k \rangle}{4} = \frac{L}{2}
\end{align}
Therefore, the statistics of motif count at the end of process in the network is always doomed to fail for the SI process.

Furthermore, we use Mahalanobis distance to address the contributions of individual submotifs to the detection of the endogeneity of the process  and present the results in Fig. \ref{fig:SI-results-c}  and \ref{fig:SI-results-d}. The square of the Mahalanobis distance follows a chi-squared distribution, with the dimension of the submotifs being equal to the degrees of freedom. Thus, the Mahalanobis distance is in general greater or equal than the z-score of a given motif. For one-edge statistic it is equal to the absolute value of the two sided z-score, and it increases as the number of submotifs increases. In general we find that the one-edge z-score is a better statistic for low percentage of defaults in a network and for lower average degree, while two-edge Mahalanobis vastly outperforms for larger percentages of defaults and higher average degrees of the network. As we have shown previously the number of causal edges converge for SI model to the random limit, but the pattern of higher order causal motifs does not, and this is the case for using Mahalanobis distance for large defaults of the network. As before in Fig.\ref{fig:SI-results-f}, one can see that the increase of the network size enables the detection of cascades even for extreme $\zeta$ values.

\subsection*{Voter model}

In the previous subsection we concluded that the properties of the \ER network and SI process make causal motif test statistics from both original and randomized process converge to the same limit. Therefore, in addition to the SI variant of the endogenous component of the process, we consider the voter model variant, which introduces scaling the number of incoming defaulted vertices with the number of all incoming vertices per vertex. It adds inhomogeneity in the defaulting process which allows the test statistic for the original process to have a different limit than for the random process, making the two more distinguishable.

KS-test results in Fig. \ref{fig:KS_SI} equivocally state that distribution of causal one-edge motifs in the case of endogenous voter model process significantly differs from purely exogenous distribution for broader range of parameters than the higher order motifs. In particular one edge statistics is superior for smaller percentages of defaults. Since the causal motif distributions do not converge to the same 100\% default limit, as in the case of SI process, one-edge dominates for all the simulated mean degrees $\langle k \rangle$.

Z-score results in Fig. \ref{fig:VMa} and Fig. \ref{fig:VMb} show a similar picture as for the SI endogenous component, except that for larger values of $\zeta$ the results become less significant faster. For $\zeta>1$ the voter model type of an endogenous process is harder to detect in the data compared to the SI type. 

We conclude that the three-edge test statistic proves most robust at lower percentages of default for highly endogenous processes, while for all the other processes one-edge is the most successful test statistic, if we use the z-score.

 The reason for this is again that the variance of the higher order motifs grows faster than the separation between the centers of the distributions. 
 
We inspect the Mahalanobis distance and show the results of the comparison in Fig. \ref{fig:VMc}. We see that for $\zeta = 1$ one-edge z-score statistic is better at lower percentages of default, while two-edge Mahalanobis statistic takes over for higher percentages, just like in the case of the SI type process. Larger values of $\zeta$ do not show significant results. 

The reason why the voter model process in the example above is harder to detect than SI is depicted in Fig.\ref{fig:VMd}. For the same values of $\alpha$ and $\beta$, less vertices defaulted endogenously for the VM model than for the SI model. Furthermore, because $\beta$ is the rate of transition per individual edge connected to a defaulted vertex, and $\alpha$ the rate of transition of each individual vertex, the actual ratio $n_{\alpha}/n_{\beta}$ is in general different than the corresponding $\zeta$, with their relation depending on the other parameters of the process and the network (Supplement). Towards the end of the process, for an SI process the value of the ratio is approximately $n_{\alpha}/n_{\beta}\approx 2$ and, for a VM process $n_{\alpha}/n_{\beta}\approx 3$, for the parameters reported in Fig.\ref{fig:VMd}. Therefore in the case of $\zeta=1$, the number of actual causal defaults in the network is 2-3 times lower than the defaults related to field effect.

\subsection*{ Defaults of Croatian companies}

As we are trying to detect default cascades, we filter only the firms that can both have their own default caused endogenously, and spread the default into the network, that is, they have to be both debtors and creditors. Under that condition we are left with 549 firms, forming a network with 1198 edges (debts), a mean degree of 4.36 (mean out and in degree are 2.18), and maximal degree 60. Our method is then applied to the temporal network and the results are shown in Fig. \ref{fig:CroDef}

Results for the completely defaulted network are shown in the form of histograms like in Fig \ref{fig:CroDefa}, where the red line represents the one-edge causal statistic count on the real process, and the blue histogram is the null distribution created by RRM. Also the temporal evolution of the default process is shown in Fig. \ref{fig:CroDefb} for the example of three-edge causal motifs. 
We used all the available statistics and in Figs \ref{fig:CroDefc},\ref{fig:CroDefd},\ref{fig:CroDefe} and \ref{fig:CroDeff}, we show the significance of the scores for different default percentages in the network of companies. The method points to possible default cascade happening in the early interval when $10-15\%$ of the companies in data set defaulted and, much more significantly, to the later period when $70-85\%$ of the network defaulted. 

 We can conclude that an endogenous propagation of defaults is borderline significant in this regions and this gives us confidence that a detailed analysis of balance sheets of companies as well as interviews with executives of companies that defaulted in that time period could reveal that there was a strong endogenous component of the defaults.  

\section*{Discussion}

Our aim was, given a set of data, to be able to determine from it whether the explanation of the observed default process requires the endogenous, network structure, or it is enough to model the dynamics as subject only to a field effect. 
We have developed a methodology which infers, from the data represented as a temporal network, whether there was a significant probability of a endogenous propagation of the default (contagion) or not, and we extensively tested limits of applicability of the method.

We tested the methodology on synthetic data, and were able to distinguish whether an endogenous component was present in the simulated process up to the value $\zeta \approx 1$, \emph{for small networks of 1000 vertices}.   Using KS-test we have detected cases in which the distributions of statistics counts were significantly different and used z-score and Malanobis distance to distinguish between purely exogenous and exogenous-endogenous cases. In borderline cases, which are generally around $\zeta \approx 1$ we show that power to distinguish processes depends not only on $\zeta$ but also on percentage of defaults, average degree and size of the network. The border between distinguishable and non-distinguishable range of parameter is complex and different for z-score and Mahalanobis distance. For large enough networks it is possible to distinguish processes even for a very large $\zeta$. It is also important to stress that Mahalanobis in general performs either equivalent or better to z-score, in the case of SI, while z-score outperforms Mahalanobis distance in the case of VM.

Finally, we applied our method on the pre-bankruptcy settlement data of Croatian companies. Based on our analysis we can conclude that there \emph{probably} was an endogenous propagation of defaults (prebankruptcy settlements) in the network, which is in agreement with previous research \cite{Zlatic} which used a more extensive dataset including the values of debts, assets of companies etc. Unfortunately default cascades in economy are very different from other more studied cascade processes, like meme propagation, where it is easy to see who copied the meme from whom. The ground truth in economy necessary involves detective work of checking the numerous court filings, financial reports of the companies and interviews with employees. Our method points to which companies such endeavour should be focused to prove without any doubt that an endogenous cascade was really present. 

We demonstrated that we can distinguish potential endogenous spreading even \emph{when using a minimal possible information}, which was the aim of this paper.

A natural extension for the future work with respect to economic cascades would be adding the amounts of debt as weights on edges, as it is reasonable to assume that larger debts are more probable to serve for propagation of default than smaller debts. Also, the next refinement of the method could be a limit on the time that passes between the default of a debtor and the default of a creditor, since the probability that the debtor's default was the cause of the creditor's default decays with time that passes. Systematic inclusion of more and more information to distinguish between mechanisms of endogenous propagation up to a point of purely data driven mechanisms is a research direction for which the presented research is a good starting point. 

In this paper we also did not enter into the question of different network classes like for instance correlated or scale-free networks \cite{newman2006structure}. Understanding the difference of detection in different classes would surely be of interest, especially since it is well known that they can have significant effects on contact processes \cite{newman2006structure,pastor2001epidemic,d2012robustness}. 

Furthermore as it is seen in this paper, sometime it is impossible to recognize whether the endogenous process exists if the "noise" of exogenous process is too strong. There should exist a fundamental phase transition between the detectable and the undetectable phase of epidemics, similar to the community detection detectability limit \cite{nadakuditi2012graph}. An analytical understanding of this limit is still challenging and would be of great importance for the researchers interested in the exogenous - endogenous interplay of interacting particle systems on complex networks.

\section*{Methods}
\subsection*{Rationale}
We devise our methodology to show the existence of a contact process component in a network process. In order to be able to observe a contact process, in our data consisting of lenders and borrowers there must exist borrowers which are also lenders. Then we can assume two possible scenarios. In the first one, the default of that vertex is completely related to exogenous causes and the connection of the lender and its borrower is irrelevant for the default of the lender - perhaps the credit was too small to significantly influence the financial stability of the lender, or there exists an institution that will repay the borrower's obligations to the lender. In the second scenario the default of the borrower at time $t_1$ \emph{causes} the default of at least one of its lenders. This lender is then present in the data as defaulted at the later time $t_2$. One can observe that in the first scenario the default of the lender and borrower can happen in any order, while in the second the order is clear $t_1$ has to happen before $t_2$.

Notice that the unknown default mechanisms for any of the cases \emph{can not} change this very basic causal relationship. Whichever the true mechanism that causes vertices in the data to default, \emph{ we expect} that the pattern of timestamps on the network in which defaults are driven by a purely exogenous process is very different from the timestamp pattern generated by the default process which has an endogenous component. 
More precisely, the purely exogenous process should not depend on the timestamps that are present in the data, because in principle any vertex could have defaulted at any time compared to any other vertex. That means that permuting the timestamps of defaults creates a set of default events just as likely as the original set.

\subsection*{Counting causal motifs}
Based on this observation we count the motifs \cite{milo2002network} with a purely causal structure on the network to verify if their count can be explained by a solely exogenous process. A directed edge pointing from $i$ to $j$ is associated with timestamps of defaults $t_i$ and $t_j$ of the vertices and we call this edge a \emph{causal edge} if $t_i<t_j$. We also define a \emph{causal motif} as any motif with \emph{all} causal edges. Examples of such causal motifs are presented in Table 1 of supplementary information. Our assumption is that such causal motifs are less common if the process is purely exogenous as compared with the mixed case in which both exogenous and endogenous dynamics exist. Therefore, our null-hypothesis states that the frequency of observed causal motifs is completely consistent with the purely exogenous default mechanism. 

\emph{Ordered motifs} \cite{paulau2015motif} were previously introduced in the case of food webs, and, unlike causal motifs, allow edges in the opposite direction of vertex ordering.
Another recent related work uses \emph{process motifs} \cite{schwarze2020motifs} as building blocks of dynamical systems operating on network. They are defined as small graphs composed of \emph{walks} on them. Yet another usage of temporal motifs can be found in  \cite{larock2020hypa}, in which authors use similar concept to ours to detect anomalies in time series on networks.

In previous works motifs are classified by the number of vertices that comprise them. As every contact process uses edges for propagation, we choose a convention where causal motifs are ordered by the number of edges they consist of. All the causal motifs up to and including order 3 are depicted in Table 1 of supplementary information.

Our methodology is based on counting the number of causal motifs of the given kind in the process data and comparing it to the count in the permuted data. A microcanonical RRM \cite{gauvin2018randomized} is used to create the permuted data, by shuffling the times on the defaulted vertices and keeping all the other network properties constrained. A distribution of motif counts obtained from the ensemble of realizations is then compared to the distribution one obtains from the real and simulated data.

As an alternative to motifs we also investigate the size of the \emph{largest causal component} and compare it to the largest causal components drawn from RRM.  The definition of this component is equivalent to the definition of the largest weakly connected component \cite{newman2006structure} in the network in which all the edges except causal ones are deleted. While finding larger motifs than the one we investigate here is extremely tedious computational task, finding the largest component is easy. One can also think of the largest causal component as the largest causal motif one can find in the network and is therefore a limiting case for large causal motifs.

\subsection*{Measures used to compare data and RRM}
In order to verify that there exists a possibility to distinguish between RRM and data, we perform a Kolmogorov-Smirnov test, which shows whether two distributions differ significantly or not. Only if we can find a significant difference in distributions, we could expect that the difference between number of motifs generated by some individual process might be significantly different from the RRM. 

The first measure we use is a usual z-score. The distribution of motif counts is very similar to a Gaussian distribution, thus making the z-score a good candidate for comparing the motif counts in data and RRM. In rare cases, where the  number of motifs is very low, as in the beginning of the default process with extremely large $\zeta$ and a low average degree, z-score is also not really applicable. For those parameter values, KS test also does not exhibit a significant difference between the distributions.
Considering that the Z-score statistic relies on the normality of distributions, we have performed a Shapiro-Wilk test on all the created distributions to check if they significantly differ from the normal distribution. We have found that for almost all parameters distributions the assumption of normality is not rejected, except for the case of very small default percentage (5\%-10\%) and for the case of largest component which is often non-normal, where it is rejected with $p < 0.01$.

Furthermore, we separate the motifs within each order and compute the Mahalanobis distance as a generalization of the z-score, to test which motifs are better to be used for the endogenous process detection. In Supplementary Table S1, we can observe that the "train" motif (a sequence of causal edges) is the least probable to be created by chance (exogenous process) compared to other submotifs of the same order. We expect that this difference within the motifs of given order is more sensitive to the possible existence of endogenous process. 

Therefore, for each count of causal motifs we create a vector $\mathbf{C}^{(i)}$ whose components represent the count of every type of submotifs of order $i$ found in the network. The dimension of this vector is 1 for 1 edge motifs, 3 for two edge motifs and 9 for three edge motifs. After the reshuffling of times in the network, we compute Mahalanobis distance $D^{(i)}$ of the causal motifs of order $i$ given as:
\begin{align}
    D^{(i)}=\sqrt{(\mathbf{C}^{(i)}-\mathbf{\mu}^{(i)})^T\Sigma_i^{-1}(\mathbf{C}^{(i)}-\mathbf{\mu}^{(i)})},
\end{align}
where $\mathbf{\mu}^{(i)}$ is average motif vector obtained in time reshuffled networks, and $\Sigma_i$ is a covariance matrix of the vector of counts in reshuffled networks. In the case of diagonal covariance matrix, Mahalanobis distance is the square root of the sum of squares of z-scores of each  vector component (individual causal motif). This property guarantees that Mahalanobis distance is always greater or equal to each of individual z-scores, and it makes it more sensitive for the detection of endogenous propagation. The p-value of Mahalanobis distance is computed using the method presented in  \cite{elfadaly2016point}.

\subsection*{Simulating stochastic processes of defaults}
We validate the methodology on data created by simulating default processes, with predetermined endogenous and exogenous contributions, on synthetic networks. Erd\H{o}s-R\'{e}nyi graphs are used as the underlying networks. 
We set the number of vertices $N$ and the expected degrees $\langle k_{in} \rangle$ and $\langle k_{out} \rangle$ to generate an ensemble of $N_{graph}$ directed networks. On each network a process of companies' defaults, with exogenous and endogenous contributions determined by parameters $\alpha$ and $\beta$, is simulated $N_{process}$ times, and a RRM containing $N_random$ shuffled instances is created.

The simulated process is composed of two Poisson processes, used for sampling exogenous and endogenous default interevent times, with rates $\alpha$ and ${\beta}_i = x_i \beta$, respectively. The endogenous rate is defined as a value common to the entire network $\beta$, weighted by some inherent vertex property $x_i$. Event times are therefore drawn from the exponential distribution with CDF:
\begin{equation}
    F(t;\lambda) = 1 - e^{-\lambda t}, \quad t \geq 0, \quad \lambda = \alpha, x_i \beta
\end{equation}
The weight $x_i$ will be set depending on the class of contact process used; SI type of process, as the simplest propagation process has $x_i = 1, \forall i$, while voter model weighs the rate of a vertex $i$ as $x_i = 1/k_i^{in}$.

Therefore, the default times for both endogenous process components are obtained as:
\begin{align}
    t_{j(SI)}^{\beta} &= t_i^{\alpha} + \Delta t_{(i,j)}^{\beta}, \quad \text{or} \nonumber \\
    t_{j(VM)}^{\beta/k_j^{in}} &= t_i^{\alpha} + k_j^{in}\Delta t_{(i,j)}^{\beta} 
\end{align}
and the simulation is performed using the event-driven algorithm \cite{kiss2017mathematics}.

By changing the ratio $\zeta := \alpha/\beta$ we control the endogenous component of the simulated process we use for the validation of our method. We stop the process after every 5 \% of the original network's vertices default and record the size of the largest component and the count of each motif. Then, we create an ensemble of time-shuffled networks and obtain distributions of the count of motifs and the size of the largest component.

\section*{Data availability statement}
Data are available on Git-Hub which is listed in references
\cite{GIT-HUB}.

\bibliography{reference}

\section*{Acknowledgements} 
VZ and IB acknowledge partial support form QuantiXLie Centre of Excellence, a project co-financed by the Croatian Government and European Union through the European Regional Development Fund - the Competitiveness and Cohesion Operational Program (Grant KK.01.1.1.01.0004, element leader N.P.). HS and VZ had their research supported by the European Regional Development Fund under the grant KK.01.1.1.01.0009 (DATACROSS). VZ also acknowledges the Croatian Science Foundation (HrZZ) Projects No. IP–2016–6–3347 and IP–2019–4–3321. Authors also want to express gratitude to Andrea Gabrielli and Ingo Scholtes for helpful discussions.

\section*{Author contributions statement}

V.Z. and H.S. conceived the problem,  V.Z., H.S. and V.P. gathered the data, I.B. made all the simulations, All authors analysed the results. and reviewed the manuscript. 

\textbf{Corresponding author}
Correspondence to Vinko Zlati\'c

\section*{Additional information}
\subsection*{Ethics declarations}

\textbf{Competing interests}
 
The authors declare no competing interests.

\end{document}


\maketitle


\noabstract


\section{Expected frequency of the motifs}

\begin{table*}[t]

\begin{center}

\begin{tabular}{| c | p{2cm} |  c | p{2cm}| }
\hline
Causal Motif & Probability & Causal Motif & Probability \\
\hline
\hline
\begin{tikzpicture}

\tikzset{vertex/.style = {shape=circle,draw,minimum size=0.025em,fill}}
\tikzset{edge/.style = {->, > = latex}}
\node[vertex] (a) at  (0,0) [label={right:$t_1$}]{};
\node[vertex] (b) at  (0,0.5)[label={right:$t_2$}] {};
\draw [thick,->] (a) -- (b);
\end{tikzpicture}$t_1<t_2$
& $p(\mathcal{M})\frac{1}{2}\pi(T)^2$
&
\begin{tikzpicture}

\tikzset{vertex/.style = {shape=circle,draw,minimum size=0.025em,fill}}
\tikzset{edge/.style = {->, > = latex}}
\node[vertex] (a) at (0.4,0) [label={right:$t_1$}] {};
\node[vertex] (b) at (0.4,0.5) [label={right:$t_2$}]{};
\node[vertex] (c) at (0.4,1) [label={right:$t_3$}]{};
\node[vertex] (d) at (0,0.5) [label={left:$t_4$}]{};
\draw [thick,->] (a) -- (b);
\draw [thick,->] (b) -- (c);
\draw [thick,->] (d) -- (c);
\end{tikzpicture}
$t_1<t_2<t_3;\;t_4<t_3$

&
$p(\mathcal{M})\frac{1}{8}\pi(T)^4$\\
\hline

\begin{tikzpicture}

\tikzset{vertex/.style = {shape=circle,draw,minimum size=0.025em,fill}}
\tikzset{edge/.style = {->, > = latex}}
\node[vertex] (a) at  (0,0) [label={right:$t_1$}]{};
\node[vertex] (b) at  (0,0.5) [label={right:$t_2$}]{};
\node[vertex] (c) at  (0,1) [label={right:$t_3$}] {};
\draw [thick,->] (a) -- (b);
\draw [thick,->] (b) -- (c);
\end{tikzpicture}
$t_1<t_2<t_3$
& 
$p(\mathcal{M})\frac{1}{6}\pi(T)^3$ & 
\begin{tikzpicture}

\tikzset{vertex/.style = {shape=circle,draw,minimum size=0.025em,fill}}
\tikzset{edge/.style = {->, > = latex}}
\node[vertex] (e) at  (1,0) [label={left:$t_1$}]{};
\node[vertex] (f) at  (1,0.5) [label={left:$t_2$}]{};
\node[vertex] (g) at  (1,1) [label={left:$t_3$}]{};
\node[vertex] (h) at  (1.4,0.5) [label={right:$t_4$}]{};
\draw [thick,->] (e) -- (f);
\draw [thick,->] (f) -- (g);
\draw [thick,->] (e) -- (h);
\end{tikzpicture}
$t_1<t_2<t_3;\;t_1<t_4$
&

$p(\mathcal{M})\frac{1}{8}\pi(T)^4$\\
\hline
\begin{tikzpicture}

\tikzset{vertex/.style = {shape=circle,draw,minimum size=0.025em,fill}}
\tikzset{edge/.style = {->, > = latex}}
\node[vertex] (a) at  (0,0) [label={below:$t_1$}]{};
\node[vertex] (b) at  (-0.4,0.5) [label={left:$t_2$}]{};
\node[vertex] (c) at  (0.4,0.5) [label={right:$t_3$}]{};
\draw [thick,->] (a) -- (b);
\draw [thick,->] (a) -- (c);
\end{tikzpicture}
$t_1<t_2,t_3$
& $p(\mathcal{M})\frac{1}{3}\pi(T)^3$ &

\begin{tikzpicture}

\tikzset{vertex/.style = {shape=circle,draw,minimum size=0.025em,fill}}
\tikzset{edge/.style = {->, > = latex}}
\node[vertex] (a) at (0,0) [label={right:$t_1$}]{};
\node[vertex] (b) at (-0.6,0.5) [label={left:$t_2$}]{};
\node[vertex] (c) at (0,0.5) [label={$t_3$}]{};
\node[vertex] (d) at (0.6,0.5) [label={right:$t_4$}]{};
\draw [thick,->] (a) -- (b);
\draw [thick,->] (a) -- (c);
\draw [thick,->] (a) -- (d);
\end{tikzpicture}
$t_1<t_2,t_3,t_4$
& $p(\mathcal{M})\frac{1}{4}\pi(T)^4$\\
\hline
\begin{tikzpicture}
\tikzset{vertex/.style = {shape=circle,draw,minimum size=0.025em,fill}}
\tikzset{edge/.style = {->, > = latex}}
\node[vertex] (d) at  (1.4,0.5) [label={$t_3$}]{};
\node[vertex] (e) at  (1,0) [label={left:$t_1$}]{};
\node[vertex] (f) at  (1.8,0) [label={right:$t_2$}]{};
\draw [thick,->] (e) -- (d);
\draw [thick,->] (f) -- (d);
\end{tikzpicture}
$t_1,t_2<t_3$&
$p(\mathcal{M})\frac{1}{3}\pi(T)^3$ &
\begin{tikzpicture}

\tikzset{vertex/.style = {shape=circle,draw,minimum size=0.025em,fill}}
\tikzset{edge/.style = {->, > = latex}}
\node[vertex] (e) at  (1,0) [label={left:$t_1$}]{};
\node[vertex] (f) at  (1.6,0) [label={below:$t_2$}]{};
\node[vertex] (g) at  (2.2,0) [label={right:$t_3$}]{};
\node[vertex] (h) at  (1.6,0.5) [label={$t_4$}]{};
\draw [thick,->] (e) -- (h);
\draw [thick,->] (f) -- (h);
\draw [thick,->] (g) -- (h);
\end{tikzpicture}
$t_1,t_2,t_3<t_4$&
$p(\mathcal{M})\frac{1}{4}\pi(T)^4$ \\
\hline
\begin{tikzpicture}

\tikzset{vertex/.style = {shape=circle,draw,minimum size=0.025em,fill}}
\tikzset{edge/.style = {->, > = latex}}
\node[vertex] (a) at  (0,0) [label={right:$t_1$}]{};
\node[vertex] (b) at  (0,0.5) [label={right:$t_2$}]{};
\node[vertex] (c) at  (0,1) [label={right:$t_3$}]{};
\node[vertex] (d) at  (0,1.5) [label={right:$t_4$}]{};
\draw [thick,->] (a) -- (b);
\draw [thick,->] (b) -- (c);
\draw [thick,->] (c) -- (d);
\end{tikzpicture}
$t_1<t_2<t_3<t_4$& 
$p(\mathcal{M})\frac{1}{24}\pi(T)^4$ &

\begin{tikzpicture}

\tikzset{vertex/.style = {shape=circle,draw,minimum size=0.025em,fill}}
\tikzset{edge/.style = {->, > = latex}}
\node[vertex] (a) at (0,0) [label={left:$t_1$}]{};
\node[vertex] (b) at (0,1) [label={left:$t_3$}]{};
\node[vertex] (c) at (1,0) [label={right:$t_2$}]{};
\node[vertex] (d) at (1,1) [label={right:$t_4$}]{};
\draw [thick,->] (a) -- (b);
\draw [thick,->] (c) -- (b);
\draw [thick,->] (c) -- (d);

\end{tikzpicture}
$t_1,t_2<t_3;\;t_2<t_4$&
$p(\mathcal{M})\frac{5}{24}\pi(T)^4$\\
\hline
\begin{tikzpicture}

\tikzset{vertex/.style = {shape=circle,draw,minimum size=0.025em,fill}}
\tikzset{edge/.style = {->, > = latex}}
\node[vertex] (a) at  (0,0) [label={right:$t_2$}]{};
\node[vertex] (b) at  (-0.4,0.5) [label={left:$t_3$}]{};
\node[vertex] (c) at  (0.4,0.5) [label={right:$t_4$}]{};
\node[vertex] (d1) at  (0,-0.5) [label={right:$t_1$}]{};
\draw [thick,->] (a) -- (b);
\draw [thick,->] (a) -- (c);
\draw [thick,->] (d1) -- (a);
\end{tikzpicture}
$t_1<t_2<t_3,t_4$
& $p(\mathcal{M})\frac{1}{12}\pi(T)^4$ &
\begin{tikzpicture}

\tikzset{vertex/.style = {shape=circle,draw,minimum size=0.025em,fill}}
\tikzset{edge/.style = {->, > = latex}}
\node[vertex] (a) at (0,0) [label={right:$t_1$}]{};
\node[vertex] (b) at (0.5,0.5) [label={right:$t_2$}]{};
\node[vertex] (c) at (0,1) [label={right:$t_3$}]{};

\draw [thick,->] (a) -- (b);
\draw [thick,->] (a) -- (c);
\draw [thick,->] (b) -- (c);

\end{tikzpicture}
$t_1<t_2<t_3$&
$p(\mathcal{M})\frac{1}{6}\pi(T)^3$\\
\hline

\begin{tikzpicture}

\tikzset{vertex/.style = {shape=circle,draw,minimum size=0.025em,fill}}
\tikzset{edge/.style = {->, > = latex}}
\node[vertex] (d) at  (1.4,0) [label={right:$t_3$}]{};
\node[vertex] (e) at  (1,-0.5) [label={left:$t_1$}]{};
\node[vertex] (f) at  (1.8,-0.5) [label={right:$t_2$}]{};
\node[vertex] (g) at  (1.4,0.5) [label={right:$t_4$}]{};
\draw [thick,->] (e) -- (d);
\draw [thick,->] (f) -- (d);
\draw [thick,->] (d) -- (g);
\end{tikzpicture}
$t_1,t_2<t_3<t_4$
&
$p(\mathcal{M})\frac{1}{12}\pi(T)^4$ &  & \\
\hline

\hline

\end{tabular}
\caption{Table with all possible causal motifs up to order 3. Their Probability represents the probability of occurrence of such a motif if only exogenous influence exists. 
}
\label{tbl:motifs}
\end{center}
      
\end{table*}

 Since our method revolves around using simulations to perturb timestamps in data, it is instructive to first consider what frequency of the motifs would one expect if only the exogenous process changed the state of vertices in the network. In that scenario, the probability of finding each individual casual motif $\mathcal{CM}$ would be proportional to \emph{(i)} probability 
$p(\mathcal{M})$ of finding the structural motif $\mathcal{M}$ on which a causal motif could develop; \emph{(ii)} probability $\mathcal{P}(T)$ that all the vertices of the given motif $\mathcal{M}$ have changed their state (got infected, have defaulted) by the time T; and \emph{(iii)} the probability that the vertices in the motif $\mathcal{M}$ have defaulted in such an order that all the edges of the motif have become causal.

In this exercise, the third probability is of interest to us. It can be computed using simple combinatorics, because only the mutual ordering of timestamps is of importance, and not any other details of the probability distribution of those timestamps. For example, for the third motif in the leftmost column of table, all that is important is that $t_1$ is the earliest timestamp out of the three timestamps and that occurs in $1/3$ of the random realizations.

In order to compute that probability for higher order motifs than the one presented here, we found that integrating the exogenous probability over time is a simpler approach. Let us suppose that the exogenous probability per unit time of the change of state of an individual vertex is $\phi(t)$. Then we can write an equation for the probability of a random appearance of the three-edge causal motif that looks like letter N in Table S\ref{tbl:motifs} as:
\begin{widetext}
\begin{align}
    P(\mathcal{CM})&=p(\mathcal{M})\int_0^T \int_0^{t_3}\left(\int_0^{t_3}\phi(t_1)dt_1\int_{t_2}^T\phi(t_4)dt_4\right)\phi(t_2)\phi(t_3)dt_2 dt_3\nonumber\\
    &=p(\mathcal{M})\int_0^T\int_0^{t_3}\pi(t_3)[\pi(T)-\pi(t_2)]\phi(t_2)dt_2\phi(t_3)dt_3\nonumber\\
    &=p(\mathcal{M})\frac{5}{24}\pi(T)^4,
\end{align}
\label{eq:integrals}
\end{widetext}

where we defined $\pi(\tau) := \int_0^\tau\phi(t)dt$, so the relation between cumulative probability and probability density is $\int \phi(t)dt\equiv \int d\pi(t)$.

As we see in the Table S\ref{tbl:motifs}, the higher the order of a motif is, the less probable it is to be found in a purely exogenous process. We will use the count for each motif order  $\mathcal{N}(t)$ at time $t$ as the statistic to test if the data can be explained only through exogenous process, by comparing it against the RRM. 

\section{The number of vertices defaulted by the exogenous and the endogenous process components}

In order to show the difference between $\zeta$ that we used as a control parameter of the processes and the actual number of vertices that defaulted through the endogenous or the exogenous process, we write rate equations using the mean field approximation.
Using those equations we compute how many vertices $n_{\alpha}$ default through the exogenous component of the process and how many vertices $n_{\beta}$ default through the endogenous process in an \ER network with $N$ vertices, and average in- and out-degrees $\langle {k}_i \rangle $ and $\langle {k}_o \rangle $.

The increase of the number of exogenously defaulted vertices, $dn_{\alpha}$, is proportional to the number of the non-defaulted vertices $N-n_{\alpha}-n_{\beta}$, rate $\alpha$ and time increment $dt$
\begin{align}
    dn_{\alpha}&=(N-n_{\alpha}-n_{\beta})\alpha dt.
\label{eq:n_a}
\end{align}
The increase of the number of endogenously defaulted vertices $dn_{\beta}$ is proportional to the number of defaulted vertices $n=n_{\alpha}+n_{\beta}$, and their average out degree $\langle {k}_o \rangle$ , which gives the number of edges through which the default can propagate. For the propagation to occur, the vertices at the ends of those edges have to be previously non-defaulted, which can be modeled with the mean field probability that the vertex is not defaulted $1-\frac{n_{\alpha}+n_{\beta}}{N}$. Together with the rate $\beta$ and time increment $dt$, all these factors lead to the equation: 
\begin{align}
dn_{\beta}&=(n_{\beta}+n_{\alpha})\langle{k}_o\rangle(1-\frac{n_{\alpha}+n_{\beta}}{N})\beta dt.
\label{eq:n_b}
\end{align}
From the equations (\ref{eq:n_a}) and (\ref{eq:n_b}), one can easily extract the time dependence in order to obtain the "phase space" differential equation, using the fact that both $n_\alpha$ and $n_{\beta}$ are monotonously increasing functions of time.

\begin{align}
\frac{dn_{\alpha}}{dn_{\beta}}&=\frac{N\zeta}{\langle k\rangle_o(n_{\alpha}+n_{\beta})}\\
dn_{\beta}&=\frac{\langle k\rangle_o}{N\zeta}(n_{\alpha}dn_{\alpha}+n_{\beta}dn_{\alpha})\label{Eq:Linear}
\end{align}
The equation (\ref{Eq:Linear}) is a linear equation that can easily be integrated to obtain:
\begin{align}
    n&=\frac{N\zeta}{\langle k\rangle_o}\left(e^{\frac{\langle k\rangle_on_{\alpha}}{N\zeta}}-1\right)\\
    \frac{n_{\alpha}}{n_{\beta}}&=\frac{\frac{\zeta}{\langle k\rangle_o}ln(\frac{d\langle k\rangle_o}{\zeta}+1)}{d-\frac{\zeta}{\langle k\rangle_o}ln(\frac{d\langle k\rangle_o}{\zeta}+1)},\label{eq:SI}
\end{align}
where $n=n_{\alpha}+n_{\beta}$ is a total number of defaulted vertices in the network and $d = n/N$ is the total default percentage in the network with $N$ vertices. For the case of VM, we substitute $\zeta\rightarrow\zeta \langle {k}_i \rangle $, and, since $\langle {k}_i \rangle =\langle {k}_o \rangle $, the equations are
\begin{align}
    n&=N\zeta\left(e^{\frac{n_{\alpha}}{N\zeta}}-1\right)\\
    \frac{n_{\alpha}}{n_{\beta}}&=\frac{\zeta ln(\frac{d}{\zeta}+1)}{d-\zeta ln(\frac{d}{\zeta}+1)},\label{eq:VM}
\end{align}

From equations (\ref{eq:SI}) and (\ref{eq:VM}) it is clear that for a given $\zeta$ and for $\langle {k}_o \rangle>1$, the VM process will always have a higher ratio $n_{\alpha}/n_{\beta}$ than the SI process. It is also clear that this ratio is monotonously decreasing with respect to the total number of defaults $d$.

\newpage
\section{Kolmogorov-Smirnov test results}

\begin{figure*}[ht!]
\centering
\begin{subfigure}{.35\textwidth}
  \centering
  \includegraphics[width=\linewidth]{supplement/ks/ks,largest_zeta=0.1,SI.eps} 
  \caption{$\zeta=0.1$,SI}
  \label{fig:KS-LCc}
\end{subfigure}%
\begin{subfigure}{.35\textwidth}
  \centering
  \includegraphics[width=\linewidth]{supplement/ks/ks,largest_zeta=0.1,VM.eps}  
  \caption{$\zeta=0.1$,VM}
  \label{fig:KS-LCd}
\end{subfigure}%
\\[1ex]
\begin{subfigure}{.35\textwidth}
  \centering
  \includegraphics[width=\linewidth]{supplement/ks/ks,largest_zeta=1.0,SI.eps}  
  \caption{$\zeta=1.0$,SI}
  \label{fig:KS-LCe}
\end{subfigure}%
\begin{subfigure}{.35\textwidth}
  \centering
  \includegraphics[width=\linewidth]{supplement/ks/ks,largest_zeta=1.0,VM.eps}  
  \caption{$\zeta=1.0$,VM}
  \label{fig:KS-LCf}
\end{subfigure}
\begin{subfigure}{.35\textwidth}
  \centering
  \includegraphics[width=\linewidth]{supplement/ks/ks,largest_zeta=10.0,SI.eps}  
  \caption{$\zeta=10.0$,SI}
  \label{fig:KS-LCg}
\end{subfigure}%
\begin{subfigure}{.35\textwidth}
  \centering
  \includegraphics[width=\linewidth]{supplement/ks/ks,largest_zeta=10.0,VM.eps}  
  \caption{$\zeta=10.0$,VM}
  \label{fig:KS-LCh}
\end{subfigure}
\caption{Kolmogorov-Smirnov test results for the largest component statistic. Next to the value of the statistic its p-value is written in brackets. Subfigures \textbf{a)}, \textbf{c)} and \textbf{e)} show the results for the SI process, and subfigures \textbf{b)}, \textbf{d)} and \textbf{f)} show the results for the voter model process.}
\label{fig:KS-LC}
\end{figure*}

\newpage

\begin{figure*}[ht!]
\centering
\begin{subfigure}{.35\textwidth}
  \centering
  \includegraphics[width=\linewidth]{supplement/ks/ks,onepath_zeta=0.1,SI.eps} 
  \caption{$\zeta=0.1$,SI}
  \label{fig:KS-OPc}
\end{subfigure}%
\begin{subfigure}{.35\textwidth}
  \centering
  \includegraphics[width=\linewidth]{supplement/ks/ks,onepath_zeta=0.1,VM.eps}  
  \caption{$\zeta=0.1$,VM}
  \label{fig:KS-OPd}
\end{subfigure}%
\\[1ex]
\begin{subfigure}{.35\textwidth}
  \centering
  \includegraphics[width=\linewidth]{supplement/ks/ks,onepath_zeta=1.0,SI.eps}  
  \caption{$\zeta=1.0$,SI}
  \label{fig:KS-OPe}
\end{subfigure}%
\begin{subfigure}{.35\textwidth}
  \centering
  \includegraphics[width=\linewidth]{supplement/ks/ks,onepath_zeta=1.0,VM.eps}  
  \caption{$\zeta=1.0$,VM}
  \label{fig:KS-OPf}
\end{subfigure}
\begin{subfigure}{.35\textwidth}
  \centering
  \includegraphics[width=\linewidth]{supplement/ks/ks,onepath_zeta=10.0,SI.eps}  
  \caption{$\zeta=10.0$,SI}
  \label{fig:KS-OPg}
\end{subfigure}%
\begin{subfigure}{.35\textwidth}
  \centering
  \includegraphics[width=\linewidth]{supplement/ks/ks,onepath_zeta=10.0,VM.eps}  
  \caption{$\zeta=10.0$,VM}
  \label{fig:KS-OPh}
\end{subfigure}
\caption{Kolmogorov-Smirnov test results for the one-edge statistic. Next to the value of the statistic its p-value is written in brackets. Subfigures \textbf{a)}, \textbf{c)} and \textbf{e)} show the results for the SI process, and subfigures \textbf{b)}, \textbf{d)} and \textbf{f)} show the results for the voter model process.}
\label{fig:KS-OP}
\end{figure*}

\newpage

\begin{figure*}[ht!]
\centering
\begin{subfigure}{.35\textwidth}
  \centering
  \includegraphics[width=\linewidth]{supplement/ks/ks,twopath_zeta=0.1,SI.eps} 
  \caption{$\zeta=0.1$,SI}
  \label{fig:KS-TPc}
\end{subfigure}%
\begin{subfigure}{.35\textwidth}
  \centering
  \includegraphics[width=\linewidth]{supplement/ks/ks,twopath_zeta=0.1,VM.eps}  
  \caption{$\zeta=0.1$,VM}
  \label{fig:KS-TPd}
\end{subfigure}%
\\[1ex]
\begin{subfigure}{.35\textwidth}
  \centering
  \includegraphics[width=\linewidth]{supplement/ks/ks,twopath_zeta=1.0,SI.eps}  
  \caption{$\zeta=1.0$,SI}
  \label{fig:KS-TPe}
\end{subfigure}%
\begin{subfigure}{.35\textwidth}
  \centering
  \includegraphics[width=\linewidth]{supplement/ks/ks,twopath_zeta=1.0,VM.eps}  
  \caption{$\zeta=1.0$,VM}
  \label{fig:KS-TPf}
\end{subfigure}
\begin{subfigure}{.35\textwidth}
  \centering
  \includegraphics[width=\linewidth]{supplement/ks/ks,twopath_zeta=10.0,SI.eps}  
  \caption{$\zeta=10.0$,SI}
  \label{fig:KS-TPg}
\end{subfigure}%
\begin{subfigure}{.35\textwidth}
  \centering
  \includegraphics[width=\linewidth]{supplement/ks/ks,twopath_zeta=10.0,VM.eps}  
  \caption{$\zeta=10.0$,VM}
  \label{fig:KS-TPh}
\end{subfigure}
\caption{Kolmogorov-Smirnov test results for the two-edge statistic. Next to the value of the statistic its p-value is written in brackets. Subfigures \textbf{a)}, \textbf{c)} and \textbf{e)} show the results for the SI process, and subfigures \textbf{b)}, \textbf{d)} and \textbf{f)} show the results for the voter model process.}
\label{fig:KS-TP}
\end{figure*}

\newpage

\begin{figure*}[ht!]
\centering
\begin{subfigure}{.35\textwidth}
  \centering
  \includegraphics[width=\linewidth]{supplement/ks/ks,threepath_zeta=0.1,SI.eps} 
  \caption{$\zeta=0.1$,SI}
  \label{fig:KS-THPc}
\end{subfigure}%
\begin{subfigure}{.35\textwidth}
  \centering
  \includegraphics[width=\linewidth]{supplement/ks/ks,threepath_zeta=0.1,VM.eps}  
  \caption{$\zeta=0.1$,VM}
  \label{fig:KS-THPd}
\end{subfigure}%
\\[1ex]
\begin{subfigure}{.35\textwidth}
  \centering
  \includegraphics[width=\linewidth]{supplement/ks/ks,threepath_zeta=1.0,SI.eps}  
  \caption{$\zeta=1.0$,SI}
  \label{fig:KS-THPe}
\end{subfigure}%
\begin{subfigure}{.35\textwidth}
  \centering
  \includegraphics[width=\linewidth]{supplement/ks/ks,threepath_zeta=1.0,VM.eps}  
  \caption{$\zeta=1.0$,VM}
  \label{fig:KS-THPf}
\end{subfigure}
\begin{subfigure}{.35\textwidth}
  \centering
  \includegraphics[width=\linewidth]{supplement/ks/ks,threepath_zeta=10.0,SI.eps}  
  \caption{$\zeta=10.0$,SI}
  \label{fig:KS-THPg}
\end{subfigure}%
\begin{subfigure}{.35\textwidth}
  \centering
  \includegraphics[width=\linewidth]{supplement/ks/ks,threepath_zeta=10.0,VM.eps}  
  \caption{$\zeta=10.0$,VM}
  \label{fig:KS-THPh}
\end{subfigure}
\caption{Kolmogorov-Smirnov test results for the three-edge statistic. Next to the value of the statistic its p-value is written in brackets. Subfigures \textbf{a)}, \textbf{c)} and \textbf{e)} show the results for the SI process, and subfigures \textbf{b)}, \textbf{d)} and \textbf{f)} show the results for the voter model process.}
\label{fig:KS-THP}
\end{figure*}

\newpage

\section{Z-score}

\begin{figure*}[ht!]
\centering
\begin{subfigure}{.35\textwidth}
  \centering
  \includegraphics[width=\linewidth]{supplement/zscore/zscore,largest_zeta=0.1,SI.eps} 
  \caption{$\zeta=0.1$,SI}
  \label{fig:Z-LCc}
\end{subfigure}%
\begin{subfigure}{.35\textwidth}
  \centering
  \includegraphics[width=\linewidth]{supplement/zscore/zscore,largest_zeta=0.1,VM.eps}  
  \caption{$\zeta=0.1$,VM}
  \label{fig:Z-LCd}
\end{subfigure}%
\\[1ex]
\begin{subfigure}{.35\textwidth}
  \centering
  \includegraphics[width=\linewidth]{supplement/zscore/zscore,largest_zeta=1.0,SI.eps}  
  \caption{$\zeta=1.0$,SI}
  \label{fig:Z-LCe}
\end{subfigure}%
\begin{subfigure}{.35\textwidth}
  \centering
  \includegraphics[width=\linewidth]{supplement/zscore/zscore,largest_zeta=1.0,VM.eps}  
  \caption{$\zeta=1.0$,VM}
  \label{fig:Z-LCf}
\end{subfigure}
\begin{subfigure}{.35\textwidth}
  \centering
  \includegraphics[width=\linewidth]{supplement/zscore/zscore,largest_zeta=10.0,SI.eps}  
  \caption{$\zeta=10.0$,SI}
  \label{fig:Z-LCg}
\end{subfigure}%
\begin{subfigure}{.35\textwidth}
  \centering
  \includegraphics[width=\linewidth]{supplement/zscore/zscore,largest_zeta=10.0,VM.eps}  
  \caption{$\zeta=10.0$,VM}
  \label{fig:Z-LCh}
\end{subfigure}
\caption{Mean values of Z-scores for the largest component statistic. Subfigures \textbf{a)}, \textbf{c)} and \textbf{e)} show the results for the SI process, and subfigures \textbf{b)}, \textbf{d)} and \textbf{f)} show the results for the voter model process.}
\label{fig:Z-LC}
\end{figure*}

\newpage

\begin{figure*}[ht!]
\centering
\begin{subfigure}{.35\textwidth}
  \centering
  \includegraphics[width=\linewidth]{supplement/zscore/zscore,onepath_zeta=0.1,SI.eps} 
  \caption{$\zeta=0.1$,SI}
  \label{fig:Z-OPc}
\end{subfigure}%
\begin{subfigure}{.35\textwidth}
  \centering
  \includegraphics[width=\linewidth]{supplement/zscore/zscore,onepath_zeta=0.1,VM.eps}  
  \caption{$\zeta=0.1$,VM}
  \label{fig:Z-OPd}
\end{subfigure}%
\\[1ex]
\begin{subfigure}{.35\textwidth}
  \centering
  \includegraphics[width=\linewidth]{supplement/zscore/zscore,onepath_zeta=1.0,SI.eps}  
  \caption{$\zeta=1.0$,SI}
  \label{fig:Z-OP-e}
\end{subfigure}%
\begin{subfigure}{.35\textwidth}
  \centering
  \includegraphics[width=\linewidth]{supplement/zscore/zscore,onepath_zeta=1.0,VM.eps}  
  \caption{$\zeta=1.0$,VM}
  \label{fig:Z-OPf}
\end{subfigure}
\begin{subfigure}{.35\textwidth}
  \centering
  \includegraphics[width=\linewidth]{supplement/zscore/zscore,onepath_zeta=10.0,SI.eps}  
  \caption{$\zeta=10.0$,SI}
  \label{fig:Z-OPg}
\end{subfigure}%
\begin{subfigure}{.35\textwidth}
  \centering
  \includegraphics[width=\linewidth]{supplement/zscore/zscore,onepath_zeta=10.0,VM.eps}  
  \caption{$\zeta=10.0$,VM}
  \label{fig:Z-OPh}
\end{subfigure}
\caption{Mean values of Z-scores for the one-edge statistic. Subfigures \textbf{a)}, \textbf{c)} and \textbf{e)} show the results for the SI process, and subfigures \textbf{b)}, \textbf{d)} and \textbf{f)} show the results for the voter model process.}
\label{fig:Z-OP}
\end{figure*}

\newpage

\begin{figure*}[ht!]
\centering
\begin{subfigure}{.35\textwidth}
  \centering
  \includegraphics[width=\linewidth]{supplement/zscore/zscore,twopath_zeta=0.1,SI.eps} 
  \caption{$\zeta=0.1$,SI}
  \label{fig:Z-TPc}
\end{subfigure}%
\begin{subfigure}{.35\textwidth}
  \centering
  \includegraphics[width=\linewidth]{supplement/zscore/zscore,twopath_zeta=0.1,VM.eps}  
  \caption{$\zeta=0.1$,VM}
  \label{fig:Z-TPd}
\end{subfigure}%
\\[1ex]
\begin{subfigure}{.35\textwidth}
  \centering
  \includegraphics[width=\linewidth]{supplement/zscore/zscore,twopath_zeta=1.0,SI.eps}  
  \caption{$\zeta=1.0$,SI}
  \label{fig:Z-TP-e}
\end{subfigure}%
\begin{subfigure}{.35\textwidth}
  \centering
  \includegraphics[width=\linewidth]{supplement/zscore/zscore,twopath_zeta=1.0,VM.eps}  
  \caption{$\zeta=1.0$,VM}
  \label{fig:Z-TPf}
\end{subfigure}
\begin{subfigure}{.35\textwidth}
  \centering
  \includegraphics[width=\linewidth]{supplement/zscore/zscore,twopath_zeta=10.0,SI.eps}  
  \caption{$\zeta=10.0$,SI}
  \label{fig:Z-TPg}
\end{subfigure}%
\begin{subfigure}{.35\textwidth}
  \centering
  \includegraphics[width=\linewidth]{supplement/zscore/zscore,twopath_zeta=10.0,VM.eps}  
  \caption{$\zeta=10.0$,VM}
  \label{fig:Z-TPh}
\end{subfigure}
\caption{Mean values of Z-scores for the two-edge statistic. Subfigures \textbf{a)}, \textbf{c)} and \textbf{e)} show the results for the SI process, and subfigures \textbf{b)}, \textbf{d)} and \textbf{f)} show the results for the voter model process.}
\end{figure*}

\newpage

\begin{figure*}[ht!]
\centering
\begin{subfigure}{.35\textwidth}
  \centering
  \includegraphics[width=\linewidth]{supplement/zscore/zscore,threepath_zeta=0.1,SI.eps} 
  \caption{$\zeta=0.1$,SI}
  \label{fig:Z-THPc}
\end{subfigure}%
\begin{subfigure}{.35\textwidth}
  \centering
  \includegraphics[width=\linewidth]{supplement/zscore/zscore,threepath_zeta=0.1,VM.eps}  
  \caption{$\zeta=0.1$,VM}
  \label{fig:Z-THPd}
\end{subfigure}%
\\[1ex]
\begin{subfigure}{.35\textwidth}
  \centering
  \includegraphics[width=\linewidth]{supplement/zscore/zscore,threepath_zeta=1.0,SI.eps}  
  \caption{$\zeta=1.0$,SI}
  \label{fig:Z-THPe}
\end{subfigure}%
\begin{subfigure}{.35\textwidth}
  \centering
  \includegraphics[width=\linewidth]{supplement/zscore/zscore,threepath_zeta=1.0,VM.eps}  
  \caption{$\zeta=1.0$,VM}
  \label{fig:Z-THPf}
\end{subfigure}
\begin{subfigure}{.35\textwidth}
  \centering
  \includegraphics[width=\linewidth]{supplement/zscore/zscore,threepath_zeta=10.0,SI.eps}  
  \caption{$\zeta=10.0$,SI}
  \label{fig:Z-THPg}
\end{subfigure}%
\begin{subfigure}{.35\textwidth}
  \centering
  \includegraphics[width=\linewidth]{supplement/zscore/zscore,threepath_zeta=10.0,VM.eps}  
  \caption{$\zeta=10.0$,VM}
  \label{fig:Z-THPh}
\end{subfigure}
\caption{Mean values of Z-scores for the three-edge statistic. Subfigures \textbf{a)}, \textbf{c)} and \textbf{e)} show the results for the SI process, and subfigures \textbf{b)}, \textbf{d)} and \textbf{f)} show the results for the voter model process.}
\label{fig:Z-TP}
\end{figure*}

\section{Mahalanobis distance}

\begin{figure*}[ht!]
\centering
\begin{subfigure}{.35\textwidth}
  \centering
  \includegraphics[width=\linewidth]{supplement/mah/mah_one-edge,zeta=0.1,dyn=SI.eps} 
  \caption{$\zeta=0.1$,SI}
  \label{fig:M-OPc}
\end{subfigure}%
\begin{subfigure}{.35\textwidth}
  \centering
  \includegraphics[width=\linewidth]{supplement/mah/mah_one-edge,zeta=0.1,dyn=VM.eps}  
  \caption{$\zeta=0.1$,VM}
  \label{fig:M-OPd}
\end{subfigure}%
\\[1ex]
\begin{subfigure}{.35\textwidth}
  \centering
  \includegraphics[width=\linewidth]{supplement/mah/mah_one-edge,zeta=1.0,dyn=SI.eps}  
  \caption{$\zeta=1.0$,SI}
  \label{fig:M-OP-e}
\end{subfigure}%
\begin{subfigure}{.35\textwidth}
  \centering
  \includegraphics[width=\linewidth]{supplement/mah/mah_one-edge,zeta=1.0,dyn=VM.eps}  
  \caption{$\zeta=1.0$,VM}
  \label{fig:M-OPf}
\end{subfigure}
\begin{subfigure}{.35\textwidth}
  \centering
  \includegraphics[width=\linewidth]{supplement/mah/mah_one-edge,zeta=10.0,dyn=SI.eps}  
  \caption{$\zeta=10.0$,SI}
  \label{fig:M-OPg}
\end{subfigure}%
\begin{subfigure}{.35\textwidth}
  \centering
  \includegraphics[width=\linewidth]{supplement/mah/mah_one-edge,zeta=10.0,dyn=VM.eps}  
  \caption{$\zeta=10.0$,VM}
  \label{fig:M-OPh}
\end{subfigure}
\caption{Mean values of Mahalanobis distances for the one-edge statistic. Subfigures \textbf{a)}, \textbf{c)} and \textbf{e)} show the results for the SI process, and subfigures \textbf{b)}, \textbf{d)} and \textbf{f)} show the results for the voter model process.}
\label{fig:M-OP}
\end{figure*}

\newpage

\begin{figure*}[ht!]
\centering
\begin{subfigure}{.35\textwidth}
  \centering
  \includegraphics[width=\linewidth]{supplement/mah/mah_two-edge,zeta=0.1,dyn=SI.eps} 
  \caption{$\zeta=0.1$,SI}
  \label{fig:M-TPc}
\end{subfigure}%
\begin{subfigure}{.35\textwidth}
  \centering
  \includegraphics[width=\linewidth]{supplement/mah/mah_two-edge,zeta=0.1,dyn=VM.eps}  
  \caption{$\zeta=0.1$,VM}
  \label{fig:M-TPd}
\end{subfigure}%
\\[1ex]
\begin{subfigure}{.35\textwidth}
  \centering
  \includegraphics[width=\linewidth]{supplement/mah/mah_two-edge,zeta=1.0,dyn=SI.eps}  
  \caption{$\zeta=1.0$,SI}
  \label{fig:M-TP-e}
\end{subfigure}%
\begin{subfigure}{.35\textwidth}
  \centering
  \includegraphics[width=\linewidth]{supplement/mah/mah_two-edge,zeta=1.0,dyn=VM.eps}  
  \caption{$\zeta=1.0$,VM}
  \label{fig:M-TPf}
\end{subfigure}
\begin{subfigure}{.35\textwidth}
  \centering
  \includegraphics[width=\linewidth]{supplement/mah/mah_two-edge,zeta=10.0,dyn=SI.eps}  
  \caption{$\zeta=10.0$,SI}
  \label{fig:M-TPg}
\end{subfigure}%
\begin{subfigure}{.35\textwidth}
  \centering
  \includegraphics[width=\linewidth]{supplement/mah/mah_two-edge,zeta=10.0,dyn=VM.eps}  
  \caption{$\zeta=10.0$,VM}
  \label{fig:M-TPh}
\end{subfigure}
\caption{Mean values of Mahalanobis distances for the two-edge statistic. Subfigures \textbf{a)}, \textbf{c)} and \textbf{e)} show the results for the SI process, and subfigures \textbf{b)}, \textbf{d)} and \textbf{f)} show the results for the voter model process.}
\label{fig:M-TP}
\end{figure*}


